\documentclass[10pt,a4paper]{article}
\usepackage[margin=23mm]{geometry}
\usepackage[T1]{fontenc}
\usepackage{lmodern,microtype,amsmath,amssymb,amsthm,mathtools}
\usepackage{enumitem,booktabs,xcolor,hyperref,needspace}
\hypersetup{colorlinks=true,linkcolor=blue!50!black,citecolor=blue!50!black,urlcolor=blue!50!black}
\hypersetup{
 pdftitle={Step Recursion: Exact Membership Spectra, Mixed Strides, and Collapse},
 pdfauthor={Kirill Osipov},
 pdfsubject={Minimal-basis step recursion, exact membership spectra, indexed intersections and collapse},
 pdfkeywords={step recursion, membership spectra, mixed strides, generalized inverses, function algebras, collapse}
}

\newtheorem{theorem}{Theorem}[section]
\newtheorem{lemma}[theorem]{Lemma}
\newtheorem{proposition}[theorem]{Proposition}
\newtheorem{corollary}[theorem]{Corollary}
\newtheorem{definition}[theorem]{Definition}
\newtheorem{remark}[theorem]{Remark}
\newcommand{\N}{\mathbb N}
\newcommand{\U}{\mathcal U}
\newcommand{\B}{\mathcal B^{\mathrm{prev}}}
\newcommand{\E}{\mathcal E^0}
\newcommand{\Sel}{\operatorname{Sel}}
\newcommand{\ZV}{\operatorname{ZV}}
\newcommand{\lcmop}{\operatorname{lcm}}
\title{Step Recursion: Exact Membership Spectra,\\Mixed Strides, and Collapse}
\author{Kirill Osipov}
\date{}
\begin{document}
\maketitle
\begin{abstract}
For a strictly increasing $G:\mathbb N\to\mathbb N$ with $G(0)\ge1$,
we study minimal-basis function algebras generated by recursion along
generalized inverses of its iterates. Bounded recursion requires an
earlier pointwise majorant; unbounded recursion has no such requirement.
If the orbit $G^{[t]}(0)$ has no polynomial upper bound, we obtain exact
fixed-function membership spectra, reverse-divisibility inclusion, and
indexed intersections that need not be single-stride algebras. Every
finite divisibility-downset is realized in both regimes, and mixed
strides admit fixed functions expressing any finite monotone Boolean
condition on their additive stride monoid. For every generator, the
realized bounded spectra form a bounded meet-semilattice: a construction
sharing a bounding input coordinate realizes finite intersections
without an unrestricted selector. Polynomial orbit growth is equivalent
to definability of each of ordinary predecessor, unary doubling, and
addition in any fixed unbounded stride algebra. In this case the algebra
is ordinary primitive recursion relative to its inverse map, giving
full collapse for primitive-recursive $G$. Ordinary predecessor belongs
to a bounded stride algebra exactly when the orbit is polynomially
bounded and $G$ has bounded successive differences; equivalently,
ordinary bounded recursion embeds. A locality obstruction explains the
additional bounded requirement, and two primitive-recursive generators
with quadratic orbits have opposite bounded collapse behavior.
\end{abstract}

\section{Introduction and main results}
We study functions generated from zero, successor, and projections by
recursion along generalized inverses of an increasing map. Two choices
matter: which strides the descent permits, and whether each recursion
requires an earlier pointwise majorant. Our results classify exact
membership spectra in the non-polynomial-orbit regime and identify
the conditions for recovering ordinary recursion.

The growth boundary is detected by elementary arithmetic. In the
unbounded system, predecessor, doubling, and addition are definable
exactly when the orbit has a polynomial bound. Bounded predecessor
additionally requires bounded generator jumps, even among generators
with quadratic orbits. Independently of growth, bounded spectra are
closed under finite intersections: sharing the bounding coordinates
of two functions gives a common majorant while preserving both as
specializations.

Let $G:\N\to\N$ be strictly increasing with $G(0)\ge1$, let
$Y_t=G^{[t]}(0)$, and let $\rho_r$ be the generalized inverse of
$G^{[r]}$. We write $\B_{G,r}$ and $\U_{G,r}$ for the bounded and
unbounded single-stride algebras (defined precisely in
Section~\ref{sec:definitions}). For a fixed total finite-arity
function $f$, its membership spectra are
\[
 \Sigma_G(f)=\{r\ge1:f\in\U_{G,r}\},\qquad
 \Sigma_G^{\mathrm{prev}}(f)=\{r\ge1:f\in\B_{G,r}\}.
\]
A divisibility-downset is a set closed under taking positive divisors.
Our main growth dichotomy is
\[
 \begin{aligned}
 (\mathrm{PB})&:\quad \exists C,k\ge1\ \forall t\quad Y_t\le C(t+1)^k,\\
 (\mathrm{NPB})&:\quad \text{no polynomial bounds }Y_t.
 \end{aligned}
\]
Let $\mathrm{PR}[h]$ denote ordinary primitive recursion relative
to a unary function $h$, and let $\E$ denote ordinary bounded
recursion from the same minimal basis, with earlier pointwise
majorants. Neither notation adds arithmetic to that basis.

For example, $G(x)=2x+1$ gives
$\rho_r(y)=\lfloor y/2^r\rfloor$ and a binary function with exact
spectrum $\{1,2,3\}$. With several available strides $Q$, another
fixed function requires both $2,3\in\langle Q\rangle$; here
$\langle Q\rangle$ is the additive monoid generated by $Q$.

\Needspace{8\baselineskip}
\begin{theorem}[Main results]\label{thm:main}
For every such $G$:
\begin{enumerate}[label=(\roman*),leftmargin=2em]
\item For each $1\le p<q$ there is a binary function $F_{p,q}$ whose
bounded and unbounded membership spectra are both
\[
 \begin{cases}
 \{r\ge1:r\mid p\text{ or }r\mid q\},&(\mathrm{NPB}),\\
 \N_{>0},&(\mathrm{PB}).
 \end{cases}
\]
\item Under $(\mathrm{NPB})$, for each nonempty finite list of strides, either family $\mathcal C_r=\U_{G,r}$
or $\mathcal C_r=\B_{G,r}$ satisfies
\[
 \mathcal C_p\subseteq\mathcal C_q\iff q\mid p,\qquad
 \bigcap_i\mathcal C_{p_i}=\mathcal C_L
       \iff L\in\{p_i\},\quad L=\lcmop(p_i).
\]
If $L$ is absent, the intersection strictly contains $\mathcal C_L$
and is no single indexed algebra. Every finite divisibility-downset,
including the empty set, is realized by the same function in both
systems; its finite arity may depend on the downset.
\item Under $(\mathrm{NPB})$, for every finite monotone Boolean
formula $\varphi$ built from atoms $[p\in\langle Q\rangle]$,
there is a fixed function $F_\varphi$ whose membership, in both
mixed bounded and mixed unbounded algebras generated by any finite
stride set $Q$, holds exactly when $\varphi$ is true.
\item Under $(\mathrm{PB})$, $\U_{G,r}=\mathrm{PR}[\rho_r]$ for all
$r$. For arbitrary $G$ and any fixed $r$, $(\mathrm{PB})$ is
equivalent to definability in $\U_{G,r}$ of each of ordinary
predecessor, unary doubling, and addition. For primitive-recursive
$G$, full collapse of these algebras is equivalent to
$(\mathrm{PB})$, with common class $\mathrm{PR}$.
\item For every $G$ and $r$, ordinary bounded recursion embeds
exactly when the orbit is polynomially bounded and $G$ has bounded
successive differences:
\[
 \E\subseteq\B_{G,r}\iff (\mathrm{PB})\ \land\
       \sup_n(G(n+1)-G(n))<\infty.
\]
Full bounded collapse requires both conditions; their sufficiency for
equality across strides is not asserted here. Under these conditions,
if also $\rho_r\in\E$, then $\B_{G,r}=\E$. Polynomial orbit
growth alone is insufficient for bounded collapse: two explicit
primitive-recursive generators with quadratic orbits have opposite
bounded behavior.
\item For every $G$, realized bounded membership spectra are closed
under finite intersections and form a bounded meet-semilattice under
inclusion. More precisely, functions with shifted-projection envelopes
admit a common finite-intersection witness for their bounded and
unbounded spectra simultaneously.
\end{enumerate}
\end{theorem}

Parts (i)--(ii) are proved in Theorems~\ref{thm:spectra}
and~\ref{thm:orders}, (iii) in Theorem~\ref{thm:boolean},
(iv)--(v) in Theorems~\ref{thm:unbounded-collapse},
\ref{thm:bounded-obstruction}, \ref{thm:examples} and
Corollary~\ref{cor:arithmetic-thresholds}, and (vi) in
Theorem~\ref{thm:bounded-meets}. \emph{Full collapse} means equality
across strides; embedding $\E$ and equality with $\E$ are separate
assertions. An \emph{indexed-nonprincipal} intersection is one unequal
to every algebra in the specified single-stride family.

\paragraph{Relation to previous work.}
The three-parameter study \cite{Osipov2026} introduced bounded
step recursion and reverse-divisibility inclusion in canonical
lower-basis sectors, varying the basis, growth row, and stride.
The mixed-stride study \cite{OsipovMixed} recovered additive monoids
of canonical descents in fixed-row classes. Here the basis stays
minimal and $G$ is arbitrary. We realize exact spectra of fixed,
possibly noncanonical functions, including finite Boolean conditions;
determine indexed intersections; and give growth and locality criteria
for ordinary recursion. Theorem~\ref{thm:bounded-meets} adds general
bounded spectral intersection closure. The arithmetic tests in
Corollary~\ref{cor:arithmetic-thresholds} are consequences of the
simulation and obstruction theorems.

Primitive iteration and unary characterizations
\cite{Axt1966,GermanoMazzanti1988,Severin2008}, bounded iteration
and recursion on notation \cite{Mazzanti2005,Mazzanti2013,Clote1999},
and lattices of honest subrecursive classes
\cite{Machtey1971,Machtey1974,Machtey1975} provide related frameworks.
Their operator and lattice results are not assumed to transfer here.

\paragraph{Proof route.}
Section~\ref{sec:definitions} gives inverse geometry and bounded
controls; Section~\ref{sec:transfer} compiles divisible strides.
Sections~\ref{sec:spectra}--\ref{sec:mixed} derive single and mixed
spectra from ancestry certificates.
Sections~\ref{sec:polynomial}--\ref{sec:bounded-collapse} establish
the collapse and embedding criteria; Section~\ref{sec:growth-examples}
gives further growth examples. Appendix~\ref{app:basis-two} treats
the separate complexity comparison for the stronger basis-two algebra.

\section{Operations, inverse geometry, and finite control}\label{sec:definitions}
Let $\N=\{0,1,\ldots\}$; $u^{[j]}$ denotes the $j$-fold iterate
of a unary map $u$. For a statement $A$, $[A]\in\{0,1\}$ denotes
its truth value and does not itself assert definability. Set
\[
 \Phi_r=G^{[r]},\quad \rho_r(0)=0,\quad
 \rho_r(y)=\min\{z:\Phi_r(z)\ge y\}\ (y>0),\quad
 D_r(y)=\min\{j:\rho_r^{[j]}(y)=0\}.
\]
The step function is $\Phi_r$; its generalized inverse $\rho_r$ is
the descent function. Strict increase over integers gives $G(x)\ge x+1$,
so $\rho_r(y)<y$ for $y>0$.

\begin{definition}[Bounded and unbounded step recursion]
Unbounded step recursion along $\rho$ maps earlier functions $a,b$
to the unique total $F$ satisfying
\begin{equation}\label{eq:sr}
 F(\mathbf x,0)=a(\mathbf x),\qquad
 F(\mathbf x,y)=b(\mathbf x,\rho(y),F(\mathbf x,\rho(y)))\quad(y>0).
\end{equation}
We write $\mathrm{USR}_\rho(a,b)=F$. Bounded step recursion
$\mathrm{SR}_\rho(a,b,M)=F$ uses the same equations and additionally
requires an earlier $M$ with $F(\mathbf x,y)\le M(\mathbf x,y)$ on
all inputs. This is a semantic admission condition, not a decision
procedure. The transition sees the predecessor address, not $y$.
\end{definition}

Starting from zero, successor $S$, and all positive-arity projections,
take finite-stage closure under composition and the indicated operation
along $\rho_r$. This gives $\U_{G,r}$ and $\B_{G,r}$, respectively.
At stage $n+1$ all data, including any required majorant, must belong
to stage $n$; preparatory stages may be inserted. Dummy fixed parameters
allow unary recursions and constants without adding zero-arity
functions. All functions have fixed positive finite arity. Ordinary
primitive recursion and ordinary bounded recursion retain their names
for the predecessor descent $\max(y-1,0)$, giving $\mathrm{PR}$ and
$\E$ respectively. ``Unbounded'' means absence of a majorant
requirement; both step recursion operations terminate.

All membership claims are extensional for a fixed $G$. Finite-table
constants may depend on finitely many values of a noncomputable $G$;
uniform extraction of their programs is not asserted. Membership of
$\rho_s$ alone does not imply closure under step recursion along it;
the simulations below establish that property when needed. Arithmetic
in semantic identities verifies the constructions and does not add
initial functions. Every displayed auxiliary majorant is global,
including on inputs outside the intended simulation.

\begin{lemma}[Inverse geometry]\label{lem:geometry}
For $r,j\ge1$,
\[
 \begin{gathered}
 \rho_r(y)\le z\iff y\le\Phi_r(z),\quad
 \rho_r^{[j]}=\rho_{rj},\quad
 \rho_r(Y_t)=Y_{\max(t-r,0)},\\
 D_r(y)\le j\iff y\le Y_{rj},\qquad
 D_r(y)=\lceil D_1(y)/r\rceil,\qquad
 \rho_r(u+c)\le\rho_r(u)+c.
 \end{gathered}
\]
The maps $\rho_r,D_r$ are nondecreasing. The gaps
$\gamma_t=Y_{t+1}-Y_t$ are positive and nondecreasing, and
\begin{equation}\label{eq:gap}
 |Y_u-Y_v|\ge\gamma_{v-1}\ge Y_v/v\quad(v\ge1, u\ne v).
\end{equation}
\end{lemma}
\begin{proof}
The adjunction is the definition of the inverse of a strictly
increasing integer map. Iterating it gives the inverse-iterate and
depth-threshold identities; counting $r$-steps gives the ceiling.
If $z=\rho_r(u)$, then $\Phi_r(z+c)\ge u+c$ for $u>0$;
at $u=0$ use $\Phi_r(c)\ge c$. This proves the last inequality
and monotonicity follows from the same thresholds. Finally $G(x)-x$
is nondecreasing, hence so is $\gamma_t$. The identity
$Y_v=\sum_{i<v}\gamma_i\le v\gamma_{v-1}$ and the neighboring
gaps prove~\eqref{eq:gap}.
\end{proof}

\begin{lemma}[Large gaps under non-polynomial orbit growth]
\label{lem:large-gaps}
Assume $(\mathrm{NPB})$. For every polynomial $P\in\N[X]$ and
every fixed integer $h\ge0$, there are arbitrarily large integers
$t>h$ such that
\[
 \gamma_{t-h-1}>P(t).
\]
In particular, no eventual polynomial bound on the orbit gaps is
possible. The assertion also applies to any nonnegative expression
which is bounded above by a fixed polynomial in $t$.
\end{lemma}
\begin{proof}
If $\gamma_{t-h-1}\le P(t)$ for all sufficiently large $t$, then,
on writing $j=t-h-1$, we obtain the eventual polynomial bound
$\gamma_j\le P(j+h+1)$. Summing this bound and absorbing the
finitely many initial gaps into a constant yields a polynomial upper
bound on $Y_n=\sum_{j<n}\gamma_j$, contrary to $(\mathrm{NPB})$.
The final assertion follows by applying the result to a polynomial
upper bound for the expression in question.
\end{proof}

\begin{lemma}[Bounded finite controls]\label{lem:controls}
Every $\B_{G,r}$ contains $\rho_r$, zero/nonzero tests, fixed
threshold tests, Boolean combinations, finite tables with constant
default, and
\[
 \ZV(v,c)=\begin{cases}0,&c=0,\\v,&c>0.\end{cases}
\]
For each fixed $M$, it also contains
$\tau_M(u)=u$ for $u\le M$ and $0$ otherwise, and the selector
$\Sel_M(a,b,c)$ between $\tau_M(a)$ and $\tau_M(b)$ according
as $c=0$ or $c>0$. The latter two functions have earlier bound $M$.
More generally, a direct conditional between earlier functions
$a(\mathbf x),b(\mathbf x)\le R(\mathbf x)$ has earlier bound $R$.
\end{lemma}
\begin{proof}
Use step recursion with base zero and transition equal to the
predecessor address to obtain $\rho_r$, bounded by the recursion
argument. Base zero and transition $1$ give the nonzero test, with
bound $1$; transition $v$ gives $\ZV$, with bound $v$.
For fixed $c$, choose $K$ with $B=Y_{rK}\ge c$. Then
\[
 [u>c]=[\rho_r^{[K]}(S^{[B-c]}u)>0].
\]
Fixed-number branches are direct step recursions with their larger
numeral as earlier bound. These and $\ZV$ supply Boolean operations,
then equality with each fixed integer. Assemble a finite table one
exceptional tuple at a time, selecting a fixed output or the earlier
table with one common constant bound. This constructs $\tau_M$;
a direct recursion on $c$ then gives $\Sel_M$ with bound $M$.
The same direct recursion proves the shared-bound assertion.
All data and bounds precede their recursions.
\end{proof}

\begin{lemma}[Common depth guards]\label{lem:guards}
Every $\B_{G,r}$ contains finite-valued $\alpha_r$ and predicates
\[
 \begin{gathered}
 D_1(y)=rD_r(y)-\alpha_r(y),\quad 0\le\alpha_r(y)<r,\\
 H_m(x,y)=[D_1(y)=D_1(x)+m]\quad(m\ge0\text{ fixed}),\\
 [D_1(x)<D_1(y)],\qquad C(x)=[x=Y_t\text{ for some }t].
 \end{gathered}
\]
\end{lemma}
\begin{proof}
Put $M=Y_r$ and define
\[
 V(0)=0,\quad V(w)=\Sel_M(z,v,v),\quad z=\rho_r(w),\ v=V(z).
\]
The earlier constant $M$ bounds it. Its first update returns zero,
the second stores the last positive address, and later updates retain
that address. Thus
\[
 E_r(y)=\Sel_M(\tau_M(y),V(y),[\rho_r(y)>0])
\]
is zero at zero and the last positive address otherwise. For $y>0$,
$1\le E_r(y)\le Y_r$ and
$D_1(y)=r(D_r(y)-1)+D_1(E_r(y))$. The finite table
$0\mapsto0$, $u\mapsto r-D_1(u)$ for $1\le u\le Y_r$
defines $\alpha_r$.

The recursion
\[
 A_r(x,0)=x,\qquad A_r(x,w)=\rho_r(A_r(x,\rho_r(w)))
\]
has earlier bound $x$ and returns $\rho_r^{[D_r(w)]}(x)$.
It tests $D_r(x)\le D_r(y)$ by $A_r(x,y)=0$, hence equality
and strict order. A fixed positive difference $D_r(x)-D_r(y)=k$
is tested by the depth-$k$ condition on $A_r(x,y)$, namely
$\rho_r^{[k]}A_r(x,y)=0$ and
$\rho_r^{[k-1]}A_r(x,y)>0$; negative differences swap inputs.
Using the finitely many values of $\alpha_r(x),\alpha_r(y)$,
the equation for $H_m$ becomes
$r(D_r(y)-D_r(x))=m+\alpha_r(y)-\alpha_r(x)$, so fixed-difference
tests and a finite Boolean table give $H_m$.

Strict $D_1$ order is lexicographic in $(D_r,-\alpha_r)$.
Finally
$C(x)=\neg H_0(x,Sx)$: the depth threshold changes between $x$
and $x+1$ precisely at the endpoints $Y_t$, including $Y_0=0$.
Every recursion just used has bound $x$ or a fixed numeral.
Arithmetic in the semantic identities is not an added primitive.
\end{proof}

\begin{lemma}[Single-coordinate envelopes]\label{lem:envelope}
Every function of a minimal-basis algebra generated by bounded step
recursion is bounded by one fixed input coordinate plus a constant;
in particular this holds in $\B_{G,r}$ and $\E$. Such a shifted projection
can be introduced before subsequent recursions. The unrestricted
selector $\Sel(a,b,c)=a$ if $c=0$, and $b$ otherwise, belongs to
every $\U_{G,r}$ but to no $\B_{G,r}$.
\end{lemma}
\begin{proof}
Induct on finite-stage construction. The basis has shifted-projection
bounds. In a composition, follow the single coordinate bounding the
head and then the coordinate bounding that argument. A new bounded
step recursion inherits the envelope of its earlier majorant. The
same induction applies to ordinary bounded recursion. The selector
is unbounded step recursion in $c$, with base $a$ and transition
$b$. No fixed coordinate bounds it up to a constant: vary $b$ when
$c>0$, or $a$ when $c=0$.
\end{proof}

\section{Exact-depth transfer of the recursion operations}\label{sec:transfer}
This section gives one simulation for both bounded and unbounded step
recursion; the bounded version records the earlier bound on every
auxiliary recursion.

\begin{lemma}[Transfer with an available exact clock]\label{lem:transfer}
Let the target minimal-basis algebra use a descent $\rho$ with
$\rho(0)=0$ and $\rho(u)<u$ for $u>0$, and let
$\sigma$ and $h$ be earlier target functions with
$\sigma(0)=0$, $\sigma(u)<u$ for $u>0$, and
\[
 D_\rho(h(y))=D_\sigma(y).
\]
Step recursion along $\sigma$ can be simulated along $\rho$ whenever
its earlier data belong to the target algebra. In a bounded target
the earlier source majorant must also belong to it; the simulation
preserves an earlier pointwise majorant.
\end{lemma}
\begin{proof}
Zero/nonzero tests and $\ZV$ have the direct bounded definitions
of Lemma~\ref{lem:controls} for any shrinking descent. First construct
\[
 I_\sigma(y,0)=y,\qquad
 I_\sigma(y,w)=\sigma(I_\sigma(y,\rho(w))).
\]
It returns $\sigma^{[D_\rho(w)]}(y)$, with earlier bound $y$ in
a bounded target. The direct conditional $J(v,c)$ returning $v$
at $c=0$ and $\sigma(v)$ otherwise has earlier bound $v$.
Next set
\[
 T(y,z,0)=y,\quad
 T(y,z,w)=J(v,[I_\sigma(v,z)>0]),\quad v=T(y,z,\rho(w)).
\]
Its state only stays or descends under $\sigma$, so its earlier
bound is $y$ for every $z,w$. If $d=D_\sigma(y)$, induction gives
\[
 T(y,z,w)=\sigma^{[\min(D_\rho(w),\max(d-D_\rho(z),0))]}(y):
\]
after $j$ descents the condition for another is $d-j>D_\rho(z)$.
Therefore
\[
 \operatorname{Tail}(y,z)=T(y,z,h(y))
 =\sigma^{[\max(d-D_\rho(z),0)]}(y).
\]
For a source function $f$ with base $a$ and transition $b$, define
\[
 \begin{aligned}
 W(\mathbf x,y,0)&=a(\mathbf x),\\
 W(\mathbf x,y,w)&=\ZV\bigl(
 b(\mathbf x,\operatorname{Tail}(y,z),W(\mathbf x,y,z)),
 [I_\sigma(y,z)>0]\bigr),\quad z=\rho(w).
 \end{aligned}
\]
For every $w$, with $t=D_\rho(w)$, its invariant is
\begin{equation}\label{eq:transfer}
 W(\mathbf x,y,w)=
 \begin{cases}f(\mathbf x,\sigma^{[d-t]}(y)),&t\le d,\\0,&t>d.
 \end{cases}
\end{equation}
At $t=0$ this is the base equation. For $1\le t\le d$, the tail
is $\sigma^{[d-t+1]}(y)$ and the gate is positive, giving exactly
the next source transition. For $t>d$ the gate is zero.

In a bounded target its earlier source majorant has, by
Lemma~\ref{lem:envelope}, an envelope $R(\mathbf x,u)=x_i+c$ or
$u+c$. Since all active source addresses are at most $y$,
$R(\mathbf x,y)$ bounds $W$ for \emph{every} $w$. It is introduced
before $W$. In an unbounded target all displayed constructions use
unbounded step recursion; no bounded membership is inferred merely
from an output estimate. Finally $W(\mathbf x,y,h(y))=f(\mathbf x,y)$.
\end{proof}

\begin{corollary}[Compilation of divisible strides]\label{cor:divisible}
For every $G$, if $r\mid p$, then
$\B_{G,p}\subseteq\B_{G,r}$ and $\U_{G,p}\subseteq\U_{G,r}$.
\end{corollary}
\begin{proof}
Write $p=kr$, $\rho=\rho_r$, $\sigma=\rho^{[k]}$. We construct
an earlier bounded target clock $h$ with $D_r(h(y))=D_p(y)$.

\emph{Retraction and bounded ascent.} Set
\[
 I(a,0)=a,\qquad I(a,w)=\rho(I(a,\rho(w)))\quad(w>0).
\]
The earlier projection $a$ bounds this recursion, and
$I(a,v)=\rho^{[D_r(v)]}(a)$. Define
\[
 \operatorname{Ret}_a(v)=I(a,I(a,v)),\qquad
 \operatorname{Up}_a(v)=I(a,\rho(I(a,\operatorname{Ret}_a(v)))).
\]
If $d=D_r(a)$ and $e=D_r(v)$, both outputs lie on the
$\rho$-orbit of $a$, have depths $\min(d,e)$ and
$\min(d,e+1)$, and are therefore at most $a$.
For completeness, applying $I(a,\cdot)$ changes depth $e$ to
$\max(d-e,0)$; applying it a second time gives $\min(d,e)$.
Starting with the orbit point at this depth, the inner $I$ in
$\operatorname{Up}$ produces depth $d-\min(d,e)$, the $\rho$
reduces that depth by one when positive, and the outer $I$
returns the point at depth $\min(d,e+1)$.
A direct conditional $\operatorname{Choose}(a,v,c)$ between
$\operatorname{Ret}_a(v)$ (at $c=0$) and
$\operatorname{Up}_a(v)$ (at $c>0$) has the earlier bound $a$.

\emph{Finite-state phase clock.} For $0\le j<k$ put
$c_j=Y_{rj}$. By Lemma~\ref{lem:controls}, the fixed finite
cyclic table $\operatorname{cyc}_k(c_j)=c_{(j+1)\bmod k}$
has a bounded extension on all inputs, with earlier constant
bound $c_{k-1}$. The $\rho$-recursion
\[
 R(0)=c_0=0,\qquad
 R(w)=\operatorname{cyc}_k(R(\rho(w)))\quad(w>0)
\]
is bounded by that earlier constant and satisfies
$R(w)=c_{D_r(w)\bmod k}$, by induction on $D_r(w)$.
The test $[R(z)=c_0]$ is a fixed-threshold table.
For $k=1$ the clock and cyclic table are constantly zero, and
this test is identically true. For all $k$, the test is applied
to the depth $D_r(\rho(w))=D_r(w)-1$, so the first rise occurs at
depth $1$ and thereafter exactly at $1+k,1+2k,\ldots$.
Now put
\[
 K(a,0)=0,\qquad
 K(a,w)=\operatorname{Choose}(a,K(a,\rho(w)),
                  [R(\rho(w))=c_0])\quad(w>0).
\]
Its earlier bound is the projection $a$ on \emph{all} inputs.
At target depths $1,1+k,1+2k,\ldots$ the state rises by one
$\rho$-depth, capped at $D_r(a)$; otherwise the retraction
preserves its current orbit point (each depth occurs only once
on that orbit). Consequently induction through every depth gives
\[
 D_r(K(a,w))=\min\{D_r(a),\lceil D_r(w)/k\rceil\}.
\]
Hence $h(y)=K(y,y)$ is boundedly available and
\[
 D_r(h(y))=\lceil D_r(y)/k\rceil=D_p(y).
\]
Lemma~\ref{lem:transfer} now translates each source recursion,
with its already compiled base, transition, and previous majorant.
Its all-depth invariant provides an earlier bound for every
auxiliary recursion; syntax induction yields both inclusions.
The construction includes $k=1$, $y=0$, and a partially filled
bottom block.
\end{proof}

\section{Numerical ancestry and exact membership spectra}\label{sec:spectra}
\subsection{Ancestry and cutoff bounds}
Expand all macros into the basis, composition, and step recursion.
An output's \emph{selected ancestry} starts at zero or one actual
input occurrence. Successor appends $S$; an address dependency appends
the actual $\rho_r$-orbit segment. Composition follows the same selected
argument occurrence of the head. At a transition, follow a fresh zero,
a fixed parameter, the predecessor address, or the immediately preceding
accumulator; the last choice continues through the preceding update.
Equal-valued occurrences remain distinct. A selector follows its chosen
data branch, not its control predicate.

A \emph{$K$-cutoff evaluation} first evaluates all actual arguments
of each recursion call, including its address $y$, with the same cutoff
rules. If $D_r(y)>K$, it returns a fresh zero without evaluating the
call's base, updates, or inverse orbit. Otherwise it unfolds the usual
orbit. All argument computations remain part of the evaluation. This
is an external convention used only for the following certificate.

\begin{lemma}[Cutoff resource certificate]\label{lem:cutoff}
For every finite derivation $\delta$ in $\U_{G,r}$ there are
polynomials $A_\delta,L_\delta\in\N[X]$ with the following
\emph{simultaneous cutoff invariant}. Fix an input tuple of maximum
$N$ and $K\ge1$. Use the call-by-value $K$-cutoff convention
above: a too-deep recursion call returns a fresh zero \emph{after}
its actual arguments have been evaluated, but without evaluating
its own body. Without any growth or buffer assumption:
\begin{enumerate}[label=(\alph*),leftmargin=2em,nosep]
\item Every numerical occurrence (including those in nested argument
computations) is at most $N+A_\delta(K)$.
\item The returned occurrence has an occurrence-respecting ancestry
word over $S,\rho_r$ of length at most $L_\delta(K)$,
starting at a fresh zero or one actual input occurrence.
\end{enumerate}
In particular, if
\begin{equation}\label{eq:buffer}
 D_r(N+A_\delta(K))\le K,
\end{equation}
no cutoff occurs, so the same assertions hold for the ordinary
evaluation. The selected word may depend on the evaluated inputs.
\end{lemma}
\begin{proof}
Induct simultaneously on occurrence bounds and selected ancestry in
the expanded derivation. Zero and projections have $A=L=0$;
successor has $A=L=1$. For $f=h(g_1,\ldots,g_s)$, take
\[
 A_f=A_h+\sum_iA_{g_i},\qquad L_f=L_h+\sum_iL_{g_i}.
\]
All argument evaluations are included in the first sum. Their maximum
is at most $N+\sum_iA_{g_i}(K)$, after which the head adds at most
$A_h(K)$. A selected head ancestry either starts at its own zero or
joins one actual argument occurrence, whose ancestry contributes at
most $\max_iL_{g_i}(K)$. Thus the second sum is sufficient.

For $F=\mathrm{USR}_{\rho_r}(a,b)$, set
\[
 A_F=A_a+KA_b,\qquad L_F=L_a+KL_b+K.
\]
A cut-off call contributes only a fresh-zero result; its argument
computations are already covered by the composition step. Otherwise
write its orbit as $w_h=y,\ldots,w_0=0$, where $h\le K$.
The base has all occurrences at most $N+A_a(K)$. At update $i$,
the fixed parameters and predecessor address are at most $N$ and
the preceding accumulator is at most
$N+A_a(K)+(i-1)A_b(K)$. The inductive certificate for $b$ therefore
bounds every occurrence in that update, including nested argument
computations, by $N+A_a(K)+iA_b(K)$. This proves (a).

For (b), follow the last update's selected ancestry backward. A
parameter exit joins its actual input; an address exit attaches at
most $h$ actual inverse edges. An accumulator exit moves to the
immediately preceding update. Hence the chain crosses at most $h$
transition certificates, one base certificate, and one address exit,
with length at most $L_a(K)+hL_b(K)+h$. A fresh zero, including a
nested cutoff, ends the chain. These estimates bound all occurrences
and one selected chain separately; they require no bound on evaluation
tree size. Finite structural induction makes $A_F,L_F$ polynomials.

Under~\eqref{eq:buffer}, (a) places every encountered recursion
address at depth at most $K$. Thus no cutoff occurs, and the evaluation
agrees node by node with ordinary evaluation.
\end{proof}

\begin{lemma}[Word separation under $(\mathrm{NPB})$]\label{lem:separation}
Suppose $(\mathrm{NPB})$ holds. If fixed $a,s_1,\ldots,s_k\ge0$
and $f\in\U_{G,r}$ satisfy
\[
 f(Y_{t+s_1},\ldots,Y_{t+s_k})=Y_{t-a}
 \quad\text{for every sufficiently large }t,
\]
then $r\mid a+s_i$ for some $i$.
\end{lemma}
\begin{proof}
A word containing $j$ descents and $b$ successors sends $Y_u$ into
\begin{equation}\label{eq:word}
 [Y_{\max(u-jr,0)},\ Y_{\max(u-jr,0)}+b].
\end{equation}
Indeed $S$ increases only the upper offset; $\rho_r$ replaces
the center index by $\max(u-r,0)$ and does not increase the offset,
by Lemma~\ref{lem:geometry}. This proves the interval invariant
in the actual chronological order, without rearranging the word.

Take the cutoff polynomials $A,L$ for $f$, and put
$v=t-a$, $n=t+\max s_i$, $N=Y_n$, $K=\lceil n/r\rceil+1$.
Lemma~\ref{lem:large-gaps}, with the fixed shift $a$ and a
polynomial upper bound for $A(K)+L(K)$ as a function of $t$,
allows us to choose an arbitrarily large layer with
\[
 Y_v>L(K),\qquad \gamma_{v-1}>A(K)+L(K).
\]
The gaps are nondecreasing and $n\ge v$, so
$N+A(K)\le Y_n+\gamma_n=Y_{n+1}\le Y_{rK}$.
Lemma~\ref{lem:cutoff} applies. A zero-origin word gives output
at most $L(K)<Y_v$. For an input origin,
\eqref{eq:word} and~\eqref{eq:gap} force its center to be exactly
$Y_v$: every other endpoint is farther than $L(K)$.
Thus $t+s_i-jr=t-a>0$ for some $i,j$, proving $r\mid a+s_i$.
\end{proof}

\subsection{Exact spectra and indexed intersections}
\begin{theorem}[Exact spectra of canonical witnesses]\label{thm:spectra}
Fix $p_1,\ldots,p_k\ge1$, set $a=\min_i p_i$, $s_i=p_i-a$,
and reorder coordinates so $s_1=0$. Define
\[
 B(\mathbf x)=\bigwedge_i C(x_i)\ \land\ \bigwedge_i H_{s_i}(x_1,x_i),
 \qquad F(\mathbf x)=\ZV(\rho_{p_1}(x_1),B(\mathbf x)).
\]
Its bounded and unbounded membership spectra are both
\begin{equation}\label{eq:spectrum}
 \begin{cases}
 \bigcup_i\{r:r\mid p_i\},&(\mathrm{NPB}),\\
 \N_{>0},&(\mathrm{PB}).
 \end{cases}
\end{equation}
For $k=2$, $p_1=p<q=p_2$, this is the binary witness
\[
 F_{p,q}(x,y)=
 \begin{cases}\rho_p(x),&C(x)\land C(y)\land H_{q-p}(x,y),\\
 0,&\text{otherwise},\end{cases}
\]
equally represented by $\rho_q(y)$ on the same guard.
\end{theorem}
\begin{proof}
The common guard belongs to every $\B_{G,r}$ and recognizes
$(Y_{t+s_1},\ldots,Y_{t+s_k})$. On that ray all
$\rho_{p_i}(x_i)$ equal $Y_{\max(t-a,0)}$. Thus each coordinate
gives the same total function through $\ZV$; if $r\mid p_i$,
$\rho_{p_i}=\rho_r^{[p_i/r]}$ gives a bounded definition at stride $r$.
Under $(\mathrm{NPB})$, membership at any other stride contradicts
Lemma~\ref{lem:separation}, since $a+s_i=p_i$.
Under $(\mathrm{PB})$, Lemma~\ref{lem:canonical-cap} below constructs
the common bounded function $\Theta(Y_t)=Y_{\max(t-1,0)}$.
Then $F=\ZV(\Theta^{[a]}(x_1),B(\mathbf x))$ is defined at every
bounded stride. That lemma's proof uses no spectrum theorem,
so this forward reference creates no circular dependency.
\end{proof}

\begin{theorem}[Inclusions, intersections, and finite spectra]\label{thm:orders}
Under $(\mathrm{NPB})$, either family $\mathcal C_r=\U_{G,r}$ or
$\mathcal C_r=\B_{G,r}$ has reverse-divisibility inclusion order.
For every nonempty finite list of strides, putting $L=\lcmop(p_i)$,
\[
 \bigcap_i\mathcal C_{p_i}=\mathcal C_L\iff L\in\{p_i\}.
\]
Otherwise the intersection strictly contains $\mathcal C_L$ and
is not any indexed $\mathcal C_r$. Every finite divisibility-downset,
including the empty set, is a common membership spectrum in both
systems; the full set $\N_{>0}$ is also a common spectrum.
\end{theorem}
\begin{proof}
Divisibility gives inclusions by Corollary~\ref{cor:divisible}.
Conversely $\rho_p\in\B_{G,p}$; if $\rho_p\in\U_{G,q}$,
its values $Y_{t-p}$ on $Y_t$ force $q\mid p$ by
Lemma~\ref{lem:separation}. This proves both inclusion orders.

If $L=p_j$, the stride-$L$ class is contained in all others and
is their intersection. Otherwise the witness in
Theorem~\ref{thm:spectra} is in every listed class but not the
stride-$L$ class. If the intersection equalled $\mathcal C_s$,
the inclusions $\mathcal C_L\subseteq\mathcal C_s\subseteq
\mathcal C_{p_i}$ would force $s\mid L$ and $p_i\mid s$ for
all $i$, hence $s=L$, a contradiction.

A nonempty finite divisibility-downset is the union of the divisor
sets of its maximal elements, so use the finite-family witness.
For the empty set use addition. If it belonged to $\U_{G,r}$,
choose its cutoff polynomials $A,L$ and arbitrarily large $t$ with
$Y_t>A(K)+L(K)$ and $\gamma_t>A(K)$, where
$K=\lceil t/r\rceil+1$. Lemma~\ref{lem:large-gaps} gives such
$t$: choose $\gamma_{t-1}>A(K)+L(K)$ and use
$Y_t\ge\gamma_{t-1}$ and $\gamma_t\ge\gamma_{t-1}$.
On $(Y_t,Y_t)$ the cutoff
condition holds, but the output is at most $Y_t+A(K)<2Y_t$.
Finally zero realizes the full spectrum. All arities stay fixed
within each membership comparison.
\end{proof}

\subsection{Closure of membership spectra}
\begin{proposition}[Further algebraic consequences]\label{prop:consequences}
For every $G,r$, $\B_{G,r}\subsetneq\U_{G,r}\subseteq\mathrm{PR}[\rho_r]$.
If $G$ is primitive recursive, $\U_{G,r}\subseteq\mathrm{PR}$, with strict
inclusion under $(\mathrm{NPB})$. The realized unbounded membership
spectra are closed under nonempty finite intersections.
\end{proposition}
\begin{proof}
Unbounded step recursion gives $J_1(x,y)=x+D_r(y)$ by incrementing
the state at each update. More generally, if $J_j$ is defined,
\[
 K_j(x,t,0)=x,\quad
 K_j(x,t,u)=J_j(K_j(x,t,\rho_r(u)),t)
\]
returns $x+D_r(u)D_r(t)^j$, so diagonal composition gives
$J_{j+1}(x,y)=x+D_r(y)^{j+1}$. Fixed compositions give
$x+P(D_r(y))$ for every fixed polynomial $P\in\N[X]$.
The unary $y+D_r(y)$ violates the shifted-coordinate envelope,
proving strict bounded/unbounded inclusion.

For arbitrary $G$, a computation by step recursion is ordinary
primitive recursive relative to $\rho_r$. To see this explicitly,
encode a reverse list by $\langle u,v\rangle=2^u(2v+1)$,
with empty list $0$. For $c>0$ its head is $\nu_2(c)$ and its
tail is $((c/2^{\nu_2(c)})-1)/2$, both ordinary primitive recursive;
set both decoders to zero at $c=0$. Maintain the list of values
through index $n$ by ordinary primitive recursion. At index $n+1$,
$\rho_r(n+1)\le n$; access that entry with
$n-\rho_r(n+1)$ tail operations, compute the transition, and prepend
its value. If $G$ is primitive recursive, $\rho_r(y)$ is obtained
by bounded search over $z\le y$ with $G^{[r]}(z)\ge y$.
Thus $\U_{G,r}\subseteq\mathrm{PR}$, and the addition exclusion
above makes it strict under $(\mathrm{NPB})$.

For arbitrary total $f,g$ on separate tuples, let
$P(\mathbf x,\mathbf y,c)=\Sel(f(\mathbf x),g(\mathbf y),c)$.
Then $\Sigma_G(P)=\Sigma_G(f)\cap\Sigma_G(g)$: one direction
uses the selector, the other restricts $c$ to $0$ or $1$ and
the unused tuple to zero. Iterate for finite intersections.
For bounded spectra the unrestricted selector cannot be used.
The next theorem obtains the corresponding closure by sharing the
input coordinate that supplies each function's envelope.
\end{proof}

\begin{theorem}[Finite intersections of bounded membership spectra]
\label{thm:bounded-meets}
Fix any $G$ as above. Suppose total functions
$f:\N^k\to\N$ and $g:\N^m\to\N$ have envelopes
\[
 f(\mathbf x)\le x_i+a,\qquad
 g(\mathbf y)\le y_j+b
\]
for fixed coordinates $i,j$ and constants $a,b\ge0$. There is a
fixed $(k+m)$-ary function $H$ such that
\[
 \begin{aligned}
 \Sigma_G^{\mathrm{prev}}(H)
   &=\Sigma_G^{\mathrm{prev}}(f)\cap\Sigma_G^{\mathrm{prev}}(g),\\
 \Sigma_G(H)&=\Sigma_G(f)\cap\Sigma_G(g).
 \end{aligned}
\]
The function $H$ itself has a shifted-projection envelope. Consequently,
for arbitrary total finite-arity functions, their realized bounded
membership spectra are closed under finite intersections. Under
inclusion these spectra form a bounded meet-semilattice, with bottom
$\varnothing$ and top $\N_{>0}$.
\end{theorem}
\begin{proof}
Let $\iota_i(z,\mathbf u)$ insert $z$ into coordinate $i$, with
$\mathbf u\in\N^{k-1}$ supplying the others in order; define
$\iota_j(z,\mathbf v)$ analogously. These are tuples of projections,
with empty remaining tuples when the arity is one. Put
\[
 H(z,\mathbf u,\mathbf v,c)=
 \begin{cases}
 f(\iota_i(z,\mathbf u)),&c=0,\\
 g(\iota_j(z,\mathbf v)),&c>0.
 \end{cases}
\]
Both branches have the global bound $z+C$, $C=\max(a,b)$ a fixed
numeral. If $f,g\in\B_{G,r}$, introduce this successor iterate
before applying the bounded conditional of Lemma~\ref{lem:controls}.
If $f,g\in\U_{G,r}$, use the unbounded conditional instead.
Conversely, recover $f$ from $H$ by substituting $z=x_i$,
$\mathbf u=(x_\ell)_{\ell\ne i}$, $\mathbf v=\mathbf0$, $c=0$;
recover $g$ using $c=1$ and the unused $\mathbf u$ coordinates zero.
Composition therefore gives both spectral equalities. The same bound
on $H$ allows finite iteration.

If arbitrary $f,g$ both have nonempty bounded spectra,
Lemma~\ref{lem:envelope} supplies envelopes for the fixed functions;
choose these once and apply the construction. If either spectrum is
empty, the intersection is realized by addition, excluded from every
bounded stride algebra by the envelope lemma. Zero realizes the full
spectrum and the empty intersection. This gives the asserted bounded
meet-semilattice.
\end{proof}

The proof uses only zero, successor, projections, composition,
the shifted-projection envelope property, and closure under conditionals
whose branches share a shifted-projection bound. It therefore applies
to any indexed family of numerical function algebras satisfying these
assumptions, independently of the particular descent functions.
This is a closure theorem for spectra of individual functions.
It coexists with Theorem~\ref{thm:orders}: the intersection of two
indexed algebras need not itself be an indexed algebra. The shared
coordinate is essential to the displayed construction, because the
unrestricted selector on two independent numerical inputs violates
the envelope lemma.

\section{Mixed strides and Boolean membership spectra}\label{sec:mixed}
For finite $Q\subseteq\N_{>0}$, let $\U_{G,Q}$ and $\B_{G,Q}$
be the finite-stage algebras over the same minimal basis, allowing
unbounded or earlier-majorized step recursion respectively along
all $\rho_q$ with $q\in Q$. The case $Q=\varnothing$ permits
composition only. Let
\[
 \langle Q\rangle=\left\{q_1+\cdots+q_s:s\ge0,\ q_i\in Q\right\}
\]
be the additive monoid including $0$; it is $\{0\}$ for the empty set.
An atom $A_p(Q)$ means $p\in\langle Q\rangle$.
The polynomially unbounded orbit hypothesis used in this section is
\begin{equation}\label{eq:npb}
  (\mathrm{NPB})\qquad Y_t\text{ has no polynomial upper bound in }t.
\end{equation}
The point of the theorem is its \emph{uniform membership witness}:
$F_\varphi$ is chosen before $Q$ is specified. Of course, the
\emph{program} representing that function may depend on the set of
available strides. The Boolean atoms describe membership of positive
integers in $\langle Q\rangle$, not additional Boolean primitives
supplied in the initial basis.

\begin{theorem}[Boolean spectra]\label{thm:boolean}
Assume~\eqref{eq:npb}. For every finite monotone Boolean formula
$\varphi$ in finitely many atoms $A_p$, $p\ge1$, there is a fixed
total finite-arity function $F_\varphi$ such that, for \emph{every}
finite $Q\subseteq\N_{>0}$,
\[
 F_\varphi\in\B_{G,Q}
 \quad\Longleftrightarrow\quad
 F_\varphi\in\U_{G,Q}
 \quad\Longleftrightarrow\quad
 \varphi\bigl((A_p(Q))_p\bigr)=\mathrm{true}.
\]
The construction is extensional from $G$, the clauses of $\varphi$, and
fixed finite tables. No computability condition on $G$ is imposed;
finite constants used in the proofs may depend nonuniformly on $G$.
\end{theorem}

\paragraph{Shared finite controls.}
Lemmas~\ref{lem:controls} and~\ref{lem:guards} imply that for
nonempty $Q$, all fixed threshold tests, canonical-point predicate
$C$, fixed-depth guards $H_m$, finite Boolean tables and zero-or-value
selectors are already in $\B_{G,Q}$. A single earlier $q\in Q$
provides the necessary construction stages. No numerical order,
addition, or unrestricted pairing is added to the initial basis.
Write $\operatorname{Can}(x)=C(x)$ when distinguishing the predicate
from a set of clause labels.

\begin{lemma}[Inverse and shadow geometry]\label{lem:shadow}
For $a,b\ge1$,
\[
 \rho_a\rho_b=\rho_{a+b},\qquad
 \rho_a(Y_u)=Y_{\max(u-a,0)},\qquad
 \rho_a(Y_u+d)\le Y_{\max(u-a,0)}+d.
\]
When $u\ge a$ and $d\ge1$, also
$\rho_a(Y_u+d)\ge Y_{u-a}+1$; in particular
$\rho_a(Y_u+1)=Y_{u-a}+1$.
The gaps $\gamma_t$ are positive and nondecreasing with
$\gamma_{t-1}\ge Y_t/t$ for $t\ge1$.
\end{lemma}
\begin{proof}
The inverse adjunction and monotonicity prove the first identities.
The upper bound is inverse nonexpansiveness, as in
Lemma~\ref{lem:geometry}; for the lower bound,
$G^{[a]}(Y_{u-a})=Y_u<Y_u+d$.
The gap statement is also in Lemma~\ref{lem:geometry}.
\end{proof}
\subsection{Mixed cutoff and ancestry}
Use the occurrence semantics of Section~\ref{sec:spectra}, now
retaining each actual inverse label $q\in Q$ on the selected path.
The cutoff test uses the common depth $D_1$.

\begin{lemma}[Simultaneous mixed cutoff certificate]
\label{lem:mixed-cutoff}
Fix a finite $Q$. For every finite derivation $\delta$ in
$\U_{G,Q}$ there are polynomials
$A_\delta,L_\delta\in\N[X]$ such that, for any tuple of actual
input values with maximum $N$ and every $K\ge1$, the following
three properties hold simultaneously.
\begin{enumerate}[label=(\alph*),leftmargin=2em]
\item In a \emph{$K$-cutoff evaluation}, evaluate every actual
 argument of a recursion call, including its address, before
 testing its depth. If the resulting address $y$ satisfies
 $D_1(y)>K$, return a fresh zero without evaluating the base,
 updates, or inverse orbit of that call. Otherwise perform the
 ordinary $\rho_q$-updates, applying the same convention to all
 nested calls. Every evaluated numerical occurrence, even inside
 an argument expression or a nested call, is at most
 $N+A_\delta(K)$.
\item The returned occurrence has a selected word over
 $\{S\}\cup\{\rho_q:q\in Q\}$ of length at most
 $L_\delta(K)$, beginning at a fresh zero or at \emph{one actual
 input occurrence} of $\delta$. It uses exactly the labels of
 inverse steps encountered on the selected path.
\item If $D_1(N+A_\delta(K))\le K$, then no cutoff can occur
 anywhere, and (a)--(b) hold for the ordinary evaluation.
\end{enumerate}
The bounds are uniform in the values of inputs and independent of
any orbit-growth hypothesis.
\end{lemma}
\begin{proof}
Use the simultaneous induction of Lemma~\ref{lem:cutoff}, with the
same basis and composition bounds. For a recursion with label $q$,
a surviving address satisfies $D_1(y)\le K$, hence $h=D_q(y)\le K$.
Its base has occurrence bound $N+A_a(K)$; update $i$ has bound
$N+A_a(K)+iA_b(K)$, including all nested argument evaluations.
The selected chain crosses at most $h$ transition certificates,
one base certificate, and one actual address segment of at most $h$
edges, each labelled $q$. Nested certificates retain their own labels.
Consequently the same recurrences work:
\[
 A_f=A_a+KA_b,\qquad L_f=L_a+KL_b+K.
\]
Arguments of a cut-off call are evaluated before it returns its fresh
zero, exactly as in the single-stride proof. Thus the induction covers
all occurrences while following only one labelled ancestry chain.
If $D_1(N+A_\delta(K))\le K$, every encountered address passes the
cutoff test, proving (c). For $Q=\varnothing$ only the basis and
composition cases occur.
\end{proof}

\begin{lemma}[Intervals for weighted inverse words]
\label{lem:word}
Let $W$ be any finite chronological word over $S$ and
$\{\rho_q:q\in Q\}$. Write $b(W)$ for its number of successors,
and $w(W)$ for the sum of its inverse labels (with $w(W)=0$ if
there are no inverse operations). For all $u\ge0$,
\[
 W(Y_u)\in
 [Y_{\max(u-w(W),0)},\,Y_{\max(u-w(W),0)}+b(W)].
\]
For a shadow input $Y_u+1$, there are two cases:
\[
 \begin{array}{ll}
 w(W)\le u:&
 Y_{u-w(W)}+1\le W(Y_u+1)
 \le Y_{u-w(W)}+b(W)+1,\\[2mm]
 w(W)>u:&0\le W(Y_u+1)\le b(W)+1.
 \end{array}
\]
These bounds allow successors and inverse steps in any order,
including words which reach the bottom layer exactly or overshoot.
\end{lemma}
\begin{proof}
Induct on the length of the \emph{chronological} word. A successor
increases the upper additive offset by one and preserves the lower
bound. An inverse $\rho_q$ maps the canonical center $Y_v$ to
$Y_{\max(v-q,0)}$ and does not increase a nonnegative upper
integer offset, by Lemma~\ref{lem:shadow}. Hence the ordinary
canonical-input interval follows with total inverse weight.

For the shadow input, keep the stronger positive-offset invariant
while the cumulative inverse weight $w$ has not exceeded $u$:
\[
 Y_{u-w}+1\le v\le Y_{u-w}+1+b.
\]
A successor preserves it. When applying $\rho_q$ with
$w+q\le u$, monotonicity and the strict shadow inequality give
\[
 \rho_q(Y_{u-w}+1)\ge Y_{u-w-q}+1,
\]
while nonexpansiveness gives the new upper bound with the same
$b$. This remains valid at \emph{equality} $w+q=u$, where the
lower bound is $Y_0+1=1$. If the next inverse has $w+q>u$,
nonexpansiveness instead gives
\[
 \rho_q(v)\le \rho_q(Y_{u-w})+(1+b)=1+b,
\]
because $\rho_q(Y_{u-w})=0$. All later successors can add at
most their number, and all inverse operations are nonincreasing.
Thus after the first overshoot the output stays between zero and
$1+b(W)$. A word with no overshoot satisfies the first case.
\end{proof}

\begin{lemma}[Mixed canonical/shadow separation]
\label{lem:separate}
Assume~\eqref{eq:npb}. Fix finite $Q$, nonempty finite
$P\subseteq\N_{>0}$, and a fixed integer $M\ge\max P$. If a fixed
$f\in\U_{G,Q}$ satisfies
\begin{equation}\label{eq:ray}
 f\bigl(Y_{t+M}+1,(Y_{t+p})_{p\in P}\bigr)=Y_t
 \quad\text{for every sufficiently large }t,
\end{equation}
possibly with additional \emph{fixed-numeral} coordinates,
then $p\in\langle Q\rangle$ for at least one $p\in P$.
\end{lemma}
\begin{proof}
Let $A,L$ be the certificate polynomials from
Lemma~\ref{lem:mixed-cutoff}, and set $K=t+M+2$. Let $C_0$ be the
largest additional fixed numeral, or zero if there is none. By Lemma~\ref{lem:large-gaps}, there are arbitrarily large $t$
for which
\[
 \gamma_{t-1}>A(K)+L(K)+C_0+2,
 \qquad Y_t>L(K)+C_0+2.
\]
The second inequality follows at the same $t$ from
$Y_t\ge\gamma_{t-1}$ (and increasing gaps).

Since $M\ge\max P$, the largest input, for large $t$, is
$N=Y_{t+M}+1$.
Monotonicity of the gaps gives
\[
 N+A(K)\le Y_{t+M+1}\le Y_K,
\]
so $D_1(N+A(K))\le K$. No cutoff occurs. We obtain one
actual ancestry word $W$ of length at most $L(K)$ from either
zero, a fixed numeral, the shadow input
$Y_{t+M}+1$, or one canonical input $Y_{t+p}$.
Its inverse weight is a sum of elements of $Q$, hence belongs
to $\langle Q\rangle$; its successor count is at most $L(K)$.

For a canonical origin $Y_{t+p}$, Lemma~\ref{lem:word}
places its output within $L(K)$ above the canonical center
$Y_{\max(t+p-w(W),0)}$. Since gaps are nondecreasing, the
distance from any \emph{other} canonical center to $Y_t$ is at
least $\gamma_{t-1}>L(K)$ for large $t$.
Consequently the only way to output $Y_t$ is
$t+p-w(W)=t$, giving $w(W)=p\in\langle Q\rangle$.
A zero or fixed-numeral origin gives at most a fixed numeral
plus $L(K)$, too small for the chosen $t$.

For the shadow origin, put $u=t+M$. If $w(W)\le u$, the
strong interval of Lemma~\ref{lem:word} applies. When
$u-w(W)=t$, its lower endpoint is $Y_t+1$, excluding $Y_t$;
when $u-w(W)\ne t$, the large gap separates its entire interval
from $Y_t$. If $w(W)>u$, the overshoot bound is at most
$L(K)+1<Y_t$. Thus the shadow cannot provide a hidden route
to the desired canonical target. Only a canonical input remains,
and its inverse weight is some $p\in P\cap\langle Q\rangle$.
\end{proof}

\begin{proposition}[Exact membership spectrum of a mixed inverse descent]
\label{prop:mixed-inverse}
Assume (NPB). For any $p\ge1$ and finite $Q\subseteq\N_{>0}$,
\[
 \rho_p\in\B_{G,Q}\quad\Longleftrightarrow\quad
 \rho_p\in\U_{G,Q}\quad\Longleftrightarrow\quad
 p\in\langle Q\rangle.
\]
\end{proposition}
\begin{proof}
If $p=q_1+\cdots+q_s$ with $q_i\in Q$, the inverse-adjunction
identity gives $\rho_p=\rho_{q_1}\circ\cdots\circ\rho_{q_s}$;
each $\rho_{q_i}$ has a previously available argument projection
as majorant, so composition gives bounded membership.
For the converse suppose $\rho_p\in\U_{G,Q}$ and let $A,L$ be
its mixed-cutoff polynomials. Choose arbitrarily large $t>p+1$
with $\gamma_{t-p-1}>A(t+2)+L(t+2)+2$, by
Lemma~\ref{lem:large-gaps} with $h=p$.
Set $K=t+2$. Because gaps are nondecreasing,
$Y_t+A(K)\le Y_{t+1}\le Y_K$; no cutoff occurs on input
$Y_t$. The ordinary output is $\rho_p(Y_t)=Y_{t-p}$ and is
represented by a selected word $W$ of length at most $L(K)$.
A zero-origin word is too small, since
$Y_{t-p}\ge\gamma_{t-p-1}>L(K)$. An input-origin word sends
$Y_t$ into
$[Y_{\max(t-w(W),0)},\,Y_{\max(t-w(W),0)}+L(K)]$.
Every canonical center other than $Y_{t-p}$ is separated from
$Y_{t-p}$ by at least $\gamma_{t-p-1}>L(K)$.
Consequently $w(W)=p$. Since every inverse label lies in $Q$,
its total weight belongs to $\langle Q\rangle$. The argument
also handles $Q=\varnothing$, for which the only possible
weight is zero.
\end{proof}

\subsection{Ray witnesses and Boolean closure}
\begin{lemma}[A clause with a safe common anchor]\label{lem:clause}
Fix a nonempty finite $P\subseteq\N_{>0}$ and $M=\max P$.
For each nonempty clause $C\subseteq P$ there is a total function
$W_C(z,(x_p)_{p\in P})$ which satisfies
\[
 0\le W_C(z,\mathbf x)\le z,
 \qquad
 W_C\in\B_{G,Q}\iff W_C\in\U_{G,Q}
 \iff C\cap\langle Q\rangle\ne\varnothing
\]
for all finite $Q$. More precisely, the specialization of $W_C$
to $z=Y_{t+M}+1$, $x_p=Y_{t+p}$ for $p\in C$ equals $Y_t$,
independently of coordinates outside $C$.
\end{lemma}
\begin{proof}
Define the common Boolean guard
\[
 B_C(z,\mathbf x)=\bigwedge_{p\in C}
       \bigl(\operatorname{Can}(x_p)\wedge H_{M+1-p}(x_p,z)\bigr),
 \qquad \operatorname{Can}(x)=[x=Y_j\text{ for some }j],
\]
where $\operatorname{Can}(x)=C(x)$ is the canonical-point predicate
from Lemma~\ref{lem:guards}; the clause set is denoted by $C$. Fix an arbitrary $p_0\in C$, and set
\[
 W_C(z,\mathbf x)=\ZV\bigl(\rho_{p_0}(x_{p_0}),B_C(z,\mathbf x)\bigr).
\]
On a true guard, all $x_p$ are canonical points whose indices
differ by exactly their stride differences; therefore their
$\rho_p(x_p)$ values are equal for all $p\in C$. Thus $W_C$
is independent of the chosen $p_0$ as a \emph{total numerical
function}. Moreover $D_1(x_p)\le D_1(z)-1$ implies $x_p<z$
on a true guard, while a false guard gives zero; hence
$0\le W_C\le z$ globally.

If some $p\in C\cap\langle Q\rangle$, then
$\rho_p$ is a composition of the permitted $\rho_q$ functions
by Lemma~\ref{lem:shadow}. For nonempty $Q$ the entire guard and
$\ZV$ are in $\B_{G,Q}$ by Lemma~\ref{lem:guards}; the
resulting definition computes $W_C$. Conversely, on the
specialization in the statement, $W_C=Y_t$ for all large $t$.
If $W_C\in\U_{G,Q}$, fix unused coordinates to zero and apply
Lemma~\ref{lem:separate} with label set $C$ and the retained
anchor $M\ge\max C$. This gives
$C\cap\langle Q\rangle\ne\varnothing$. If $Q$ is empty, the same
separation argument applies with words containing only successor.
\end{proof}

\begin{proof}[Proof of Theorem~\ref{thm:boolean}]
First suppose $\varphi$ is neither identically true nor
identically false. Let $P$ contain all its atoms. Write its
\emph{monotone conjunctive-normal form}:
\begin{equation}\label{eq:cnf}
 \varphi=\bigwedge_{i=1}^{m}\bigvee_{p\in S_i} A_p,
 \qquad \varnothing\ne S_i\subseteq P.
\end{equation}
The fixed clauses $S_i$ are nonempty and depend only on $\varphi$.
Let $W_i=W_{S_i}$ with common anchor $z$, and define
\[
 F_\varphi(z,(x_p)_{p\in P},c)=
 \begin{cases}
 W_i(z,(x_p)_{p\in P}),&c=i\in\{1,\ldots,m\},\\
 0,&\text{otherwise}.
 \end{cases}
\]
This is a fixed total function; each $W_i\le z$ on all inputs.
If~\eqref{eq:cnf} is true for $Q$, every clause has a witness
$p\in S_i\cap\langle Q\rangle$. Lemma~\ref{lem:clause}
constructs every $W_i$ in $\B_{G,Q}$.
Finite table tests and repeated bounded conditionals with the
\emph{previously available common majorant $z$} therefore construct
$F_\varphi$ in $\B_{G,Q}$. Conversely, if
$F_\varphi\in\U_{G,Q}$, substitute the fixed numeral $i$
for $c$; composition gives $W_i\in\U_{G,Q}$ for every $i$.
Lemma~\ref{lem:clause} makes each CNF clause true, and hence
$\varphi$ true. This proves both equivalences.

If $\varphi$ is identically true, choose the zero function,
which belongs to every mixed algebra including $Q=\varnothing$.
If $\varphi$ is identically false, choose binary addition.
Suppose that addition belongs to $\U_{G,Q}$ for some finite $Q$.
Let $A$ be its mixed cutoff occurrence polynomial. By
Lemma~\ref{lem:large-gaps}, choose an arbitrarily large $t$ with
$\gamma_{t-1}>A(t+2)+1$, and put $N=Y_t$, $K=t+2$.
Then $Y_t\ge\gamma_{t-1}>A(K)$ and
$N+A(K)\le Y_{t+1}\le Y_K$. The mixed cutoff certificate
therefore applies to the ordinary evaluation at $(Y_t,Y_t)$:
its output is at most $Y_t+A(K)<2Y_t$, contradicting the
required value of addition. This also covers $Q=\varnothing$.
Thus the constant-false case is realized. These choices complete
the proof.
\end{proof}

\subsection{Worked example: two OR clauses and one AND}
Take
$\varphi=(A_2\lor A_3)\land(A_5\lor A_7)$,
$P=\{2,3,5,7\}$ and $M=7$. Its clause guards are
\[
\begin{aligned}
 B_{23}&=\operatorname{Can}(x_2)\land\operatorname{Can}(x_3)
       \land H_6(x_2,z)\land H_5(x_3,z),\\
 B_{57}&=\operatorname{Can}(x_5)\land\operatorname{Can}(x_7)
       \land H_3(x_5,z)\land H_1(x_7,z).
\end{aligned}
\]
Let $W_{23}=\ZV(\rho_2(x_2),B_{23})$ and
$W_{57}=\ZV(\rho_5(x_5),B_{57})$, and select the former
when $c=1$, the latter when $c=2$, and zero otherwise.
Each branch is globally bounded by $z$. On the ray
$z=Y_{t+7}+1$, $x_p=Y_{t+p}$, both branches equal $Y_t$.
The separation lemma forces the two clauses individually;
for instance $Q=\{2,7\}$ satisfies them while $Q=\{2\}$ does not.

\begin{proposition}[A direct two-stride conjunction witness]\label{prop:two}
Under~\eqref{eq:npb}, the fixed binary function
\[
 J_{2,3}(x,c)=\begin{cases}\rho_2(x),&c=0,\\
 \rho_3(x),&c>0\end{cases}
\]
satisfies, for every finite $Q$,
\[
 J_{2,3}\in\B_{G,Q}\iff J_{2,3}\in\U_{G,Q}
 \iff\bigl(2\in\langle Q\rangle\ \text{and}\ 3\in\langle Q\rangle\bigr).
\]
\end{proposition}
\begin{proof}
If both strides are sums of generators from $Q$, the two
composed descents are earlier bounded functions, and a direct
conditional on $c$ has the common earlier majorant $x$.
Conversely, specializing $c$ to the fixed numerals $0$ and $1$
in any mixed unbounded definition yields $\rho_2$ and $\rho_3$.
Proposition~\ref{prop:mixed-inverse} forces
$2,3\in\langle Q\rangle$.
\end{proof}

\begin{remark}[Scope]
For noncomputable $G$, these are extensional membership claims and
not uniform algorithms which recover the finite control constants
from an index for $G$. This theorem classifies \emph{finite monotone formulas} in stride
monoid membership atoms, not every possible membership spectrum.
It does not assert an exhaustive description of mixed-algebra
inclusions, arbitrary infinite Boolean conditions, or a canonical
vertical hierarchy. No statement that a mixed algebra is generated
as a set by the union of single-stride algebras is inferred.
\end{remark}

\section{Polynomial orbits and full unbounded collapse}\label{sec:polynomial}
Assume $(\mathrm{PB})$ throughout this section. We first construct a
canonical cap without numerical order, then comparison below that cap.
Only the unbounded step recursion argument uses a larger numerical
carrier to obtain unrestricted order.

\subsection{Canonical caps and comparison}
\begin{lemma}[A bounded canonical predecessor for polynomial orbits]\label{lem:canonical-cap}
Assume $(\mathrm{PB})$. For every $r\ge1$ the function
\[
 \Theta(0)=0,\qquad
 \Theta(x)=Y_{D_1(x)-1}\quad(x>0)
\]
belongs to $\B_{G,r}$.
\end{lemma}
\begin{proof}
\emph{Clipping and capped successor.}
All controls in Lemma~\ref{lem:guards} are available.
Define the depth clip
\[
 T(x,v)=\mathrm{ZV}\bigl(v,[D_1(v)<D_1(x)]\bigr).
\]
For $x>0$, the threshold identity gives
$0\le T(x,v)\le\Theta(x)<x$; at $x=0$ it is zero.
The function is a composition of already bounded functions.
Put $q=T(x,v)$ and let
\[
 c(x,v)=C(q)\land H_1(q,x).
\]
For $x>0$, this predicate is true precisely when $q=\Theta(x)$.
A direct conditional recursion with fixed parameters $(x,v)$, base
$S(T(x,v))$, and positive branch $T(x,v)$ is admitted with the
earlier majorant $x+1$. Composing its test argument with $c(x,v)$
and applying a zero-or-value gate at $x$ gives a bounded function
\[
 Q(x,v)=\begin{cases}
 0,&x=0,\\
 q,&x>0\text{ and }c(x,v)=1,\\
 q+1,&x>0\text{ and }c(x,v)=0.
 \end{cases}
\]
For every input tuple, $Q(x,v)\le\Theta(x)\le x$: after clipping,
$q\le\Theta(x)$, and an increment occurs only when $q<\Theta(x)$.
On the intended interval $0\le v\le\Theta(x)$ it is exactly the
successor capped at $\Theta(x)$.

\emph{Polynomial iteration.} Define
\[
 R_1(x,v,0)=T(x,v),\qquad
 R_1(x,v,w)=Q\bigl(x,R_1(x,v,\rho_r(w))\bigr)\quad(w>0),
\]
and put $P_1(x,v)=R_1(x,v,x)$. Having defined $P_j$, define
\[
 R_{j+1}(x,v,0)=T(x,v),\qquad
 R_{j+1}(x,v,w)=P_j\bigl(x,R_{j+1}(x,v,\rho_r(w))\bigr)
 \quad(w>0),
\]
and $P_{j+1}(x,v)=R_{j+1}(x,v,x)$.
Every one of these recursions has the earlier projection $x$ as a
pointwise majorant for all $(x,v,w)$, since both its base and every
transition output lie between zero and $\Theta(x)$.
Writing $d=D_r(x)$, induction gives the semantic identities
\[
 P_j(x,v)=\min\{T(x,v)+d^j,\Theta(x)\}.
\]
Products, powers, and the minimum in this identity describe the
number of nested capped updates; they are not assumed generators.

Choose integers $C,k\ge1$ with $Y_t\le C(t+1)^k$ for all $t$ and
put $M=Cr^k$. For $x>0$, let $t=D_1(x)\le rd$. Then
\[
 \Theta(x)=Y_{t-1}\le Ct^k\le M d^k.
\]
Apply $P_k(x,\cdot)$ successively $M$ times to zero, using a fixed
finite composition. Its output is $\Theta(x)$; at zero it also
returns zero. Every required function and earlier majorant has
already been constructed, proving bounded membership at stride $r$.
\end{proof}

\begin{lemma}[Comparison below a canonical cap]\label{lem:interior}
Assume $(\mathrm{PB})$ and fix $r\ge1$. There is a Boolean function
$I_r(x,u,v)\in\B_{G,r}$ such that
\[
 0\le u,v\le\Theta(x)\quad\Longrightarrow\quad
 I_r(x,u,v)=[u<v].
\]
No condition is imposed on its values outside this interval.
\end{lemma}
\begin{proof}
Use $T,Q,P_j$ from Lemma~\ref{lem:canonical-cap}. Their definitions
give a clip $T(x,v)$ into $[0,\Theta(x)]$, the successor $Q(x,v)$
capped at $\Theta(x)$ on that interval, and
$P_j(x,v)=\min\{T(x,v)+d^j,\Theta(x)\}$, where $d=D_r(x)$.
The Boolean test
\[
 e(x,v)=[x>0]\land C(T(x,v))\land H_1(T(x,v),x)
\]
recognizes $v=\Theta(x)$ on the intended interval when $x>0$.
All these functions are already bounded functions.

For $i\ge0$ construct a capped shift $V_i(x,v,w)$ by
$\rho_r$-recursion, starting at $T(x,v)$ and applying $Q(x,\cdot)$
when $i=0$, or $P_i(x,\cdot)$ when $i>0$. Every such recursion has
the earlier bound $x$ on all inputs, and
\[
 V_i(x,v,w)=\min\{T(x,v)+D_r(w)d^i,\Theta(x)\}.
\]
Choose fixed $M,k\ge1$ with $\Theta(x)\le M d^k$ for $x>0$, as
in Lemma~\ref{lem:canonical-cap}. Applying $P_k$ a fixed $m$ times
and then $V_0,\ldots,V_{k-1}$ constructs capped shifts by
\[
 j=m d^k+\sum_{i=0}^{k-1}D_r(w_i)d^i,
 \qquad 0\le m<M.
\]
Apply these shifts separately to $u$ and $v$.

Existential quantification over the digits $D_r(w_i)=0,\ldots,d-1$
is an available bounded operation. Explicitly, for any earlier
Boolean predicate $A(\mathbf z,w)$ use
\[
 E(\mathbf z,0)=0,\qquad
 E(\mathbf z,w)=E(\mathbf z,\rho_r(w))\lor
                  A(\mathbf z,\rho_r(w)),
\]
with the earlier constant bound $1$, and evaluate at $w=x$.
The addresses on that orbit have depths $0,\ldots,d-1$.
Repeat this construction for the $k$ digits and take a fixed finite
disjunction over $m$. Thus all $j=0,\ldots,Md^k-1$ are tested.

Return true if for one such $j$ the capped shift of $v$ has reached
$\Theta(x)$ and the capped shift of $u$ has not. For
$u,v\le\Theta(x)$ this happens exactly when $u<v$: in that case
take $j=\Theta(x)-v$, which is less than $Md^k$ because $v\ge1$.
If $u\ge v$, it never happens. At $x=0$ return zero. All numerical
auxiliary recursions have bound $x$, and all quantifiers have bound
$1$, proving the assertion including the earlier-majorant condition.
\end{proof}

\subsection{Compiling ordinary recursion}
\begin{lemma}[Polynomial clocks simulate ordinary recursion]\label{lem:ordinary-compiler}
Assume $(\mathrm{PB})$ and fix a stride $r$.
\begin{enumerate}[label=(\roman*),leftmargin=2em]
\item If ordinary numerical order is in $\U_{G,r}$, then this
algebra generated by unbounded step recursion is closed under ordinary primitive recursion.
\item If ordinary numerical order is in
$\B_{G,r}$, then this algebra generated by bounded step recursion is closed
under ordinary bounded recursion.
\end{enumerate}
In part~(ii), closure means that the base, transition, and pointwise
majorant are already in the target algebra before the simulated
ordinary recursion is introduced.
\end{lemma}
\begin{proof}
\emph{Capped clocks.} Numerical order and finite controls give
\[
 L_y(v)=\begin{cases}v,&v\le y,\\0,&v>y,\end{cases}
 \qquad q_y(v)=\min\{L_y(v)+1,y\}.
\]
The first is a zero-or-value gate. The second selects between
$L_y(v)+1$ and $y$ with earlier bound $y+1$, and always returns at
most $y$. As in Lemma~\ref{lem:canonical-cap}, nest and iterate
this capped successor to obtain, for $d=D_r(y)$,
\[
 A_i(y,s,w)=\min\{L_y(s)+D_r(w)d^i,y\},\qquad i\ge0.
\]
For $i=0$ iterate $q_y$ along $w$; for $i>0$ first use the nesting
that constructed $P_i$, then iterate its shift by $d^i$ along $w$.
Write that shift as $p_i(y,s)$ and put $p_0=q_y$.
Every bounded clock recursion has the earlier bound $y$ on all
$s,w$; the unbounded construction uses the same equations without
majorants.

\emph{Safe updates.} Let the source equations be
\[
 f(\mathbf x,0)=a(\mathbf x),\qquad
 f(\mathbf x,n+1)=b(\mathbf x,n,f(\mathbf x,n)).
\]
In the bounded case, apply Lemma~\ref{lem:envelope} to the earlier
target majorant $M$, giving a shifted projection $R(\mathbf x,y)$ with
\[
 n\le y\quad\Longrightarrow\quad
 f(\mathbf x,n)\le M(\mathbf x,n)\le R(\mathbf x,y).
\]
Choose $R=x_i+c$ if $M\le x_i+c$, and $R=y+c$ if $M\le n+c$.
Let $\kappa_R(v)$ retain $v\le R$ and return zero otherwise;
numerical order makes this gate available. In the unbounded case
set $\kappa_R$ to the identity throughout the proof and use the
unrestricted selector. Define $B_0(\mathbf x,y,s,v)$ by
\[
 \begin{cases}
 \kappa_R\bigl(b(\mathbf x,L_y(s),\kappa_R(v))\bigr),&L_y(s)<y,\\
 \kappa_R(v),&L_y(s)=y.
 \end{cases}
\]
Both bounded branches share the earlier majorant $R$. Recursively set
\[
\begin{aligned}
 H_j(\mathbf x,y,s,v,0)&=\kappa_R(v),\\
 H_j(\mathbf x,y,s,v,w)&=
 B_j\bigl(\mathbf x,y,A_j(y,s,z),H_j(\mathbf x,y,s,v,z)\bigr),
 \quad z=\rho_r(w),\\
 B_{j+1}(\mathbf x,y,s,v)&=H_j(\mathbf x,y,s,v,y).
\end{aligned}
\]
Every bounded $H_j$ has earlier bound $R(\mathbf x,y)$ for
arbitrary $s,v,w$, since its base and each update are clipped.

\emph{Block invariant.} For verification only, define the map on
index/state pairs $0\le s\le y$ by
\[
 \mathfrak t_y(s,v)=
 \begin{cases}
 (s+1,\kappa_R(b(\mathbf x,s,\kappa_R(v)))),&s<y,\\
 (y,\kappa_R(v)),&s=y.
 \end{cases}
\]
We claim, for $y>0$ and every $v$,
\begin{equation}\label{eq:block-invariant}
 B_j(\mathbf x,y,s,v)
 =\pi_2\bigl(\mathfrak t_y^{[d^j]}(s,v)\bigr),
 \qquad d=D_r(y),\quad 0\le s\le y.
\end{equation}
For $j=0$ this is the definition. The initial clip is harmless because
$\mathfrak t_y(s,v)=\mathfrak t_y(s,\kappa_R(v))$; hence every
positive block is unchanged by preclipping its initial state. For the
induction step, at update $i$ of $H_j(\cdots,y)$ the predecessor
address $z$ has depth $i$, so
\[
 A_j(y,s,z)=\min\{s+i d^j,y\}.
\]
Inductively this is the index after $i$ blocks of $d^j$ safe updates,
and the accumulator is their exact state. The next $B_j$ performs
the next block. The $d$ updates of $H_j$ therefore establish
\eqref{eq:block-invariant} for $j+1$. The pairs are bookkeeping;
the index is supplied by $A_j$, not stored by a pairing function.

\emph{Final composition.} Choose $C,k\ge1$ with
$Y_t\le C(t+1)^k$ and put $K=C(r+1)^k$. For $y>0$,
$y\le Y_{rd}\le Kd^k$. Starting at $a(\mathbf x)$, apply $B_k$
in $K$ consecutive blocks, with starting indices
$(p_k(y,\cdot))^{[m]}(0)$, $0\le m<K$. This fixed finite
composition executes the source updates at $0,\ldots,y-1$ and
then holds. Bounded clips preserve all genuine states, which are
at most $R$. At $y=0$, the blocks return the clipped initial state;
this equals $a(\mathbf x)$ in the bounded case as well. Thus the
output is $f(\mathbf x,y)$, with all auxiliary bounds introduced
before their recursions.
\end{proof}

\begin{theorem}[Full collapse for unbounded step recursion]\label{thm:unbounded-collapse}
Under $(\mathrm{PB})$, $\U_{G,r}=\mathrm{PR}[\rho_r]$ for every $r$.
For primitive-recursive $G$, full collapse is equivalent to
$(\mathrm{PB})$, with common class $\mathrm{PR}$.
\end{theorem}
\begin{proof}
The polynomial increments in Proposition~\ref{prop:consequences}
are growth-independent constructions. Choose $C,k\ge1$ with
$Y_t\le C(t+1)^k$, and set $K=C(r+1)^k$. Fixed composition gives
\[
 N(x)=x+K D_r(x)^k+1\in\U_{G,r}.
\]
If $x>0$, $t=D_1(x)$ and $d=D_r(x)$, then
$Y_t\le Kd^k<N(x)$, so $\Theta(N(x))\ge Y_t$.
For $D_1(x)=D_1(y)$, both inputs lie below this cap. Therefore
\[
 [x<y]=[D_1(x)<D_1(y)]\lor
 \bigl([D_1(x)=D_1(y)]\land I_r(N(x),x,y)\bigr).
\]
The formula also handles $x=y=0$. Lemma~\ref{lem:ordinary-compiler}(i)
now gives ordinary primitive recursion, and $\rho_r$ is already
internal; hence $\mathrm{PR}[\rho_r]\subseteq\U_{G,r}$.
Proposition~\ref{prop:consequences} proved the reverse inclusion.
For primitive-recursive $G$, the inverse is primitive recursive
by bounded search over $z\le y$, so the common class is
$\mathrm{PR}$. Under $(\mathrm{NPB})$, reverse divisibility
precludes equality of strides $1$ and $2$.
\end{proof}

For arbitrary $G$ under $(\mathrm{PB})$, full unbounded collapse is
equivalent to $\rho_1\in\mathrm{PR}[\rho_r]$ for every $r$,
because $\rho_r=\rho_1^{[r]}$. No claim that growth alone settles
these relative definability questions is made.

\begin{corollary}[Polynomial-orbit mixed unbounded collapse]\label{cor:mixed-pb}
Assume (PB) and let $Q$ be a nonempty finite stride set. Then
\[
 \U_{G,Q}=\mathrm{PR}\bigl[(\rho_q)_{q\in Q}\bigr].
\]
If $G$ is primitive recursive, every such mixed algebra equals
ordinary $\mathrm{PR}$. This does not assert corresponding bounded
collapse or any general collapse for nonprimitive-recursive $G$.
\end{corollary}
\begin{proof}
Every $\rho_q$ belongs to $\U_{G,Q}$, and each mixed recursion
is ordinary primitive recursive relative to all the $\rho_q$,
giving one inclusion. Choose $q_0\in Q$.
Theorem~\ref{thm:unbounded-collapse} provides numerical order in
$\U_{G,q_0}$. The compiler in Lemma~\ref{lem:ordinary-compiler}
works with arbitrary earlier base and transition functions from
the larger mixed algebra, proving closure under ordinary primitive
recursion and the reverse inclusion. Primitive-recursive $G$
has primitive-recursive inverses by bounded search.
\end{proof}

\section{Bounded step recursion: locality and collapse}\label{sec:bounded-collapse}
The finite-fiber lemma assumes $(\mathrm{PB})$; the locality and
nonmembership results below it require no orbit-growth hypothesis
beyond those explicitly stated. The final embedding criterion
combines the polynomial-orbit and locality obstructions.

\subsection{Embedding and locality}
\begin{lemma}[Finite fibers and ordinary bounded recursion]\label{lem:fibers}
Under $(\mathrm{PB})$, if all fibers of $\rho_r$ have size at most
a fixed $L$, then $\E\subseteq\B_{G,r}$. If also $\rho_r\in\E$,
then $\B_{G,r}=\E$.
\end{lemma}
\begin{proof}
Put
\[
 A(a,b)=[D_1(a)<D_1(b)]\lor
 \bigl([D_1(a)=D_1(b)]\land I_r(a,\rho_r(a),\rho_r(b))\bigr).
\]
At equal positive depths both inverse values are at most
$\Theta(a)$; at $a=b=0$ the comparison also returns false.
Thus $A$ tests strict depth order, or strict inverse order at
equal depth, and is false for $a\ge b$. We have
\[
 [a<b]=\bigvee_{j=0}^{L-1}A(a+j,b+j).
\]
For if $a<b$ and every term were false, monotonicity would give
$\rho_r(a+j)=\rho_r(b+j)$ for $0\le j<L$, making $\rho_r$
constant on $[a,b+L-1]$, at least $L+1$ integers. This contradicts
the fiber bound. Numerical order is therefore available in
$\B_{G,r}$, and Lemma~\ref{lem:ordinary-compiler}(ii) gives
ordinary bounded recursion and the first inclusion.

If $\sigma=\rho_r\in\E$, ordinary iteration constructs
$J(y,n)=\sigma^{[n]}(y)$ with earlier bound $y$. The counter
\[
 C(y,0)=0,\qquad C(y,n+1)=C(y,n)+[J(y,n)>0]
\]
equals $\min(n,D_r(y))\le y$ for every $n$. Its hold/increment
transition is a direct conditional with bound $v+1$ for the
state parameter $v$, and the recursion itself has earlier bound
$y$. Hence $h(y)=C(y,y)=D_r(y)$ belongs to $\E$.
With ordinary predecessor as target descent, $h$ is an exact
source-depth clock. Lemma~\ref{lem:transfer} translates bounded
step recursion along $\sigma$ into ordinary bounded recursion,
including earlier majorants. Syntax induction proves the reverse
inclusion $\B_{G,r}\subseteq\E$.
\end{proof}

\begin{lemma}[Locality of bounded step recursion]\label{lem:locality}
For every unary $f\in\B_{G,r}$ there is $c_f\ge0$ such that,
if $\rho_r$ is constant on an integer interval $[A,B]$, $A\ge1$,
and $B-A\ge c_f$, the unfolded evaluation of a fixed derivation
of $f$ on $[A,B-c_f]$ has a uniform occurrence-labelled shape:
each corresponding numerical occurrence is either constant or
$x+j$ for a fixed $j\ge0$. In particular, the restriction of $f$
to this interval is constant or $x+j$ for one fixed $j\ge0$.
\end{lemma}
\begin{proof}
\emph{Uniform overhead.} Every bounded derivation $\delta$ has a
constant $c_\delta$ such that all occurrences on input maximum $N$
are at most $N+c_\delta$. For zero, successor, and projections take
$0,1,0$; for $h(\mathbf g)$ take $\max_i c_{g_i}+c_h$.
For a bounded recursion with data $a,b,M$, the envelope lemma gives
$M(\mathbf x,w)\le x_i+m$ or $w+m$. Every genuine accumulator
and address is at most $N+m$. The base and each transition therefore
have occurrence bounds $N+c_a$ and $N+m+c_b$, respectively.
Thus $c_F=\max\{m,c_a,m+c_b,c_M\}$ suffices. This structural
induction covers all nested evaluations, regardless of orbit length.

\emph{Parametric evaluation.} Expand the fixed unary program and
choose $c_f$ covering its all-occurrences overhead. At $x=A$, label
each occurrence by a fixed integer or $x+j$, $j\ge0$, propagating
labels from the syntax and keeping distinct occurrences separate.
Zero is fixed; projections copy; successor increments. Inverse
application to a fixed label remains fixed. If its argument is
$x+j$, the overhead at $A$ implies $j\le c_f$, so
\[
 [A+j,B-c_f+j]\subseteq[A,B].
\]
Flatness makes its inverse value fixed throughout $[A,B-c_f]$.
Labels are never inferred from accidental equalities at $A$.

Induct simultaneously on syntax and the finite updates of each call.
Composition evaluates its arguments and then its head, preserving
the assigned labels. A fixed recursion address has a fixed orbit and
zero status. A variable address $x+j$ stays positive and has the same
fixed first predecessor on the whole interval; its remaining orbit
is fixed as well. The base and successive transitions therefore have
the same call structure. Applying the constituent syntax induction
at each actual parameter and accumulator occurrence preserves the
labels through every update. This is well-founded because a transition
uses only earlier functions and the preceding accumulator.

Thus the reference evaluation at $A$ is reproduced throughout the
interval, with each corresponding occurrence fixed or $x+j$.
Majorants supply the overhead, not extra branching operations; if
evaluated, they are earlier derivations covered by the same induction.
The output label gives the claimed dichotomy.
\end{proof}

\begin{lemma}[Predecessor obstruction for non-polynomial orbits]
\label{lem:pred-npb}
If $(\mathrm{NPB})$ holds, then ordinary predecessor
$P(x)=\max(x-1,0)$ does not belong to $\U_{G,r}$ for any $r\ge1$.
Consequently $\E\nsubseteq\B_{G,r}$.
\end{lemma}
\begin{proof}
Suppose $P$ has a derivation in $\U_{G,r}$, with cutoff
polynomials $A,L$ as in Lemma~\ref{lem:cutoff}. Under (NPB),
there are arbitrarily large $t$ for which
\[
 \gamma_{t-1}>A(K)+L(K)+2,\qquad
 K=\lceil t/r\rceil+1.
\]
This follows from Lemma~\ref{lem:large-gaps}, since the right-hand
side is bounded by a fixed polynomial in $t$. Put $N=Y_t$. Since $\gamma_t\ge\gamma_{t-1}$,
\[
 N+A(K)\le Y_{t+1}\le Y_{rK},
\]
so the cutoff evaluation is the ordinary evaluation. Its output
$P(Y_t)=Y_t-1$ has a selected word of length at most $L(K)$
from zero or the input $Y_t$. From zero, it is at most $L(K)$,
strictly below $Y_t-1$. From the input, a word with no inverse
edges has value at least $Y_t$. If it contains an inverse edge,
Lemma~\ref{lem:geometry} and the chronological word interval
\eqref{eq:word} give value at most
$Y_{\max(t-r,0)}+L(K)\le Y_{t-1}+L(K)<Y_t-1$.
All possible numerical origins are excluded. The last assertion
follows because $P\in\E$ and $\B_{G,r}\subseteq\U_{G,r}$.
\end{proof}

\begin{theorem}[Unbounded forward differences prevent bounded collapse]
\label{thm:bounded-obstruction}
For every $G$, if $\sup_n(G(n+1)-G(n))=\infty$, then
\[
 \rho_p\notin\B_{G,q}\qquad(1\le p<q).
\]
Consequently full bounded collapse requires $(\mathrm{PB})$ and
bounded successive differences of $G$. For every $G$ and $r$,
\[
 \E\subseteq\B_{G,r}
 \iff (\mathrm{PB})\ \land\
        \sup_n(G(n+1)-G(n))<\infty.
\]
\end{theorem}
\begin{proof}
Suppose $\rho_p\in\B_{G,q}$ and let $c$ be its locality
constant. Write $H=G^{[q-p]}$. Increasing integer maps do not
contract distances, so $H(n+1)-H(n)\ge G(n+1)-G(n)$.
Choose $n$ with this difference greater than $c+1$, and put
\[
 A=G^{[q]}(n)+1,\qquad B=G^{[q]}(n+1).
\]
The descent $\rho_q$ is constant on $[A,B]$, whereas
\[
 \rho_p(A)=H(n)+1,\quad \rho_p(B)=H(n+1),\quad
 \rho_p(B-c)-\rho_p(A)\ge H(n+1)-H(n)-c-1>0.
\]
Here $B-c>A$ follows from the same noncontraction inequality,
and the last bound uses the inverse Lipschitz inequality.
Thus $\rho_p$ is nonconstant and strictly below the identity
on the shrunken interval, contradicting Lemma~\ref{lem:locality}.
Under $(\mathrm{NPB})$, Theorem~\ref{thm:orders} gives distinct
stride-$1$ and stride-$2$ bounded algebras. Together with the first
assertion, this proves both necessary conditions for full bounded
collapse.

When $(\mathrm{NPB})$ holds, Lemma~\ref{lem:pred-npb} already
excludes ordinary predecessor even from $\U_{G,r}$, so
$\E\nsubseteq\B_{G,r}$. This establishes the necessity
of $(\mathrm{PB})$ independently of the jump condition.
If successive differences are bounded by $L$, positive fibers
of $\rho_r$ have size at most $L^r$ and its zero fiber has
size $Y_r+1$; under $(\mathrm{PB})$, Lemma~\ref{lem:fibers}
gives $\E\subseteq\B_{G,r}$.
If successive differences are unbounded, fibers of $\rho_r$
have unbounded lengths. The locality lemma excludes ordinary
predecessor from $\B_{G,r}$, while predecessor belongs to
$\E$. These cases prove the global equivalence.
\end{proof}

\subsection{Arithmetic thresholds and examples}
\begin{corollary}[Arithmetic tests for the growth and locality thresholds]
\label{cor:arithmetic-thresholds}
Let $P(x)=\max(x-1,0)$, $\operatorname{dbl}(x)=2x$, and
$\operatorname{add}(x,y)=x+y$. For every $G$ and every $r\ge1$,
\[
 \begin{aligned}
 (\mathrm{PB})
 &\iff P\in\U_{G,r}
 \iff\operatorname{dbl}\in\U_{G,r}
 \iff\operatorname{add}\in\U_{G,r},\\
 P\in\B_{G,r}
 &\iff\E\subseteq\B_{G,r}
 \iff (\mathrm{PB})\ \land\
       \sup_n(G(n+1)-G(n))<\infty.
 \end{aligned}
\]
Thus each of the three unbounded arithmetic membership tests is
independent of the chosen stride. No computability assumption on
$G$ is required.
\end{corollary}
\begin{proof}
Under (PB), Theorem~\ref{thm:unbounded-collapse} gives
$\U_{G,r}=\mathrm{PR}[\rho_r]$, which contains all three ordinary
primitive-recursive arithmetic functions. Under (NPB),
Lemma~\ref{lem:pred-npb} excludes $P$. To exclude either doubling
or addition, take its cutoff occurrence polynomial $A$ from
Lemma~\ref{lem:cutoff}. Lemma~\ref{lem:large-gaps} supplies
arbitrarily large $t$ with
\[
 \gamma_{t-1}>A(K),\qquad K=\lceil t/r\rceil+1.
\]
Then $Y_t>A(K)$ and
$Y_t+A(K)\le Y_{t+1}\le Y_{rK}$, so no cutoff occurs on
input $Y_t$ for doubling or $(Y_t,Y_t)$ for addition. The output
is at most $Y_t+A(K)<2Y_t$, contradicting its required value.
This proves the first line.

If $P\in\B_{G,r}$, the first line and
$\B_{G,r}\subseteq\U_{G,r}$ imply (PB). Unbounded successive
differences of $G$ give arbitrarily long positive flat intervals
of $\rho_r$, since increasing integer maps do not contract
distances. On such an interval of sufficiently large length,
Lemma~\ref{lem:locality} excludes $P$: its restriction is
nonconstant and equals $x-1$, not $x+j$ for any $j\ge0$.
Thus generator differences must be bounded as well.
Theorem~\ref{thm:bounded-obstruction} then gives
$\E\subseteq\B_{G,r}$. Conversely, this embedding supplies $P$
because $P\in\E$. The same theorem gives the final equivalence.
\end{proof}

\begin{theorem}[Quadratic orbits with opposite bounded behavior]\label{thm:examples}
The primitive-recursive generators
\[
 G_+(x)=x+2\lfloor\sqrt x\rfloor+1,\qquad
 G_-(x)=x+2^{m(x)},\quad m(x)=\max\{j:4^j\le x+1\},
\]
satisfy $G_\pm(x)/x\to1$, $G_\pm(x)-x\asymp\sqrt{x+1}$,
and $G_\pm^{[t]}(0)\asymp t^2$. Nevertheless
\[
 \B_{G_+,r}=\E\quad(r\ge1),\qquad
 \B_{G_-,2}\subsetneq\B_{G_-,1}.
\]
For both generators $\U_{G_\pm,r}=\mathrm{PR}$ at every stride.
\end{theorem}
\begin{proof}
Both surpluses are nondecreasing and both maps take $0$ to $1$,
so they are strictly increasing. For $z=t^2+i$, $0\le i<2t+1$,
\begin{equation}\label{eq:square-iterate}
 G_+^{[r]}(z)=(t+r)^2+i=z+2rt+r^2.
\end{equation}
In particular its orbit is $t^2$ and its successive differences
are at most $3$. Thus fibers of $\rho_r$ have sizes bounded by
$\max(3^r,r^2+1)$.

We verify $\rho_r\in\E$ with total bounded transitions. Ordinary
predecessor is ordinary recursion bounded by its argument;
its iteration gives truncated subtraction, hence order and minimum.
Iterating successor capped at $x$, from $\min(a,x)$, gives
addition capped at $x$ with earlier bound $x$ on all inputs.
Iterating capped addition gives capped multiplication with the
same bound. For integer square root use
\[
 h(z,n,v)=\begin{cases}
 \min(n+1,z),&(n+1)^2\le z,\\
 \min(v,z),&\text{otherwise}.
 \end{cases}
\]
The predicate uses the square capped at $z+1$; both branches are
globally bounded by $z$, so direct conditional recursion defines
$h$. Ordinary bounded recursion from zero returns
$\min(w,\lfloor\sqrt z\rfloor)$ after $w$ updates; set $w=z$.
Now evaluate~\eqref{eq:square-iterate} capped at $x$ for $z\le x$,
using a fixed $2r$ capped additions and $r^2$ capped successors.
Let $P(x,z)$ test whether the result is $x$. With $\bar v=\min(v,x)$,
the update
\[
 q(x,v)=\begin{cases}\bar v,&P(x,\bar v),\\
 \min(\bar v+1,x),&\text{otherwise}
 \end{cases}
\]
has both branches bounded by the earlier projection $x$ on all
inputs. Iterate it from zero for $x$ ordinary updates. It stops at
the least satisfying $z\le x$, giving $\rho_r(x)$ also at zero.
Lemma~\ref{lem:fibers} yields $\B_{G_+,r}=\E$.

For $G_-$, bounded search computes $m(x)$ and
$\tfrac12\sqrt{x+1}<2^{m(x)}\le\sqrt{x+1}$.
Writing $u_t=\sqrt{G_-^{[t]}(0)+1}$ and rationalizing the
difference gives
\[
 \frac1{2(\sqrt2+1)}<u_{t+1}-u_t\le\frac12,
\]
so its orbit is also quadratic. Its successive differences are
unbounded at $x=4^{\ell+1}-2$, where the surplus jumps from
$2^\ell$ to $2^{\ell+1}$. Theorem~\ref{thm:bounded-obstruction}
excludes $\rho_1$ from $\B_{G_-,2}$, while it belongs to
$\B_{G_-,1}$; Corollary~\ref{cor:divisible} gives the proper
inclusion. More explicitly, if $h=2^\ell$, $\ell\ge1$,
$n=4h^2-2$, then on $[n+3h+1,n+4h+1]$ one has
$\rho_2(y)=n+1$ and $\rho_1(y)=y-2h$: the adjacent forward
second iterates are $n+3h$ and $n+4h+1$, and the intermediate
source surplus is $2h$. This also recovers the direct flat-interval
counterexample. The unbounded equalities follow from
Theorem~\ref{thm:unbounded-collapse}.
\end{proof}

\begin{corollary}[The additive endpoint]\label{cor:additive}
For $G(x)=x+1$, all strides collapse within each family:
\[
 \U_{G,r}=\mathrm{PR},\qquad \B_{G,r}=\E.
\]
Thus each membership spectrum is empty or full, and any intersection
is the least-common-multiple indexed class, even for incomparable strides.
\end{corollary}
\begin{proof}
The orbit is $Y_t=t$ and $\rho_r(y)=\max(y-r,0)$.
Theorem~\ref{thm:unbounded-collapse} applies. The fibers have
bounded sizes, and $\rho_r\in\E$ is a fixed iterate of ordinary
predecessor. Lemma~\ref{lem:fibers} proves the bounded equality.
\end{proof}

\section{Orbit-growth examples}\label{sec:growth-examples}
\begin{proposition}[Uniform bounds in the faster-growth regime]\label{prop:uniform}
Suppose
\begin{equation}\label{eq:sp}
 (\mathrm{SP}):\qquad Y_t/t^k\longrightarrow\infty
       \quad\text{for every fixed }k\ge1.
\end{equation}
Then every $f\in\U_{G,r}$ has a polynomial $P$ such that all
occurrences and the selected word length on input maximum $N$ are
at most $N+P(D_r(N)+1)$ and $P(D_r(N)+1)$, respectively.
In particular $f(\mathbf x)\le N+o(N)$; addition and unary doubling
are absent from every $\U_{G,r}$.
\end{proposition}
\begin{proof}
Let $d=D_r(N)$ and $K=d+1$. For large $N$, the gap $\gamma_{rd}$
dominates $A(K)$ by~\eqref{eq:gap} and $(\mathrm{SP})$, whence
$N+A(K)\le Y_{rd+1}\le Y_{rK}$. Apply the cutoff certificate;
finitely many smaller inputs are absorbed into constant coefficients.
Moreover $(\mathrm{SP})$ is equivalent to
$D_1(x)=o(x^\varepsilon)$ for every $\varepsilon>0$.
For one direction, if $t=D_1(x)$ then $x>Y_{t-1}$ and choose a
power $k$ with $k\varepsilon>1$. For the other, evaluate the depth
estimate at $x=Y_t$ and use $\varepsilon=1/(k+1)$.
Thus every fixed polynomial in $D_r(N)+1$ is $o(N)$. Addition
at $(N,N)$ and doubling at $N$ contradict the resulting envelope.
\end{proof}

The exact orbit condition for the canonical witnesses is $(\mathrm{NPB})$.
It is strictly weaker than eventual domination of every polynomial,
condition~\eqref{eq:sp}. For example, construct positive nondecreasing
orbit gaps as follows. At an index $a_j$, raise the gap to
$B_j\ge(a_j+2)^{j+3}$, also exceeding the previous gap, and hold it
constant until an index $b_j>a_j+1$ with $Y_{b_j}\le b_j^2$.
Such $b_j$ exists because that segment is linear. Continue from
$a_{j+1}=b_j$. Then $Y_{a_j+1}\ge(a_j+2)^{j+3}$ excludes every
polynomial upper bound, while $Y_{b_j}\le b_j^2$ excludes
\eqref{eq:sp}. The map
$G(x)=x+(Y_{i+1}-Y_i)$ for $Y_i\le x<Y_{i+1}$ is strictly
increasing and has precisely this orbit. Equivalently, $(\mathrm{NPB})$
means $\limsup_t\log(Y_t+1)/\log(t+2)=\infty$.

\begin{proposition}[Nearly linear examples]\label{prop:nearlinear}
Each of
\[
 G_2(x)=\max(1,2x),\quad
 G_\varepsilon(x)=x+1+\lfloor\varepsilon x\rfloor\ (\varepsilon>0),\quad
 G_{\log}(x)=x+1+\lfloor x/\log(x+2)\rfloor
\]
satisfies~\eqref{eq:sp}, and $G_{\log}(x)/x\to1$.
\end{proposition}
\begin{proof}
All three maps are strictly increasing and send $0$ to $1$.
For the last map, $x/\log(x+2)$ is increasing: its derivative has
the sign of $\log(x+2)-x/(x+2)>0$, whose derivative is
$x/(x+2)^2\ge0$. The first two orbits eventually grow geometrically;
indeed $G_2^{[t]}(0)=2^{t-1}$ for $t\ge1$, and
$G_\varepsilon(x)\ge(1+\varepsilon)x$.
For each integer $m\ge2$, eventually
$G_{\log}(x)\ge x+x^{1-1/m}$. The mean value theorem gives
\[
 (x+x^{1-1/m})^{1/m}-x^{1/m}\ge\frac1{2m}\qquad(x\ge1).
\]
Thus its orbit satisfies $Y_t\ge c_mt^m$ eventually. Taking $m>k$
proves $Y_t/t^k\to\infty$ for each $k$. Finally,
$(G_{\log}(x)-x)/x\le1/x+1/\log(x+2)\to0$.
\end{proof}
The raw map $2x$ fails our zero-based termination convention:
$G(0)=0$ and its inverse sends $1$ to $1$. The correction in $G_2$
is necessary. By contrast $x+1$ is valid but gives the full collapse
of Corollary~\ref{cor:additive}.

\begin{theorem}[No slowest generator]\label{thm:no-slowest}
Suppose $G$ satisfies $(\mathrm{NPB})$. Write
$\Delta(x)=G(x)-x$, $d(x)=D_1^G(x)$, and define the external generator
\[
 \eta(x)=\max_{u\le x}\left\lceil\frac{\Delta(u)}{d(u)+1}\right\rceil,
 \qquad H(x)=x+\eta(x).
\]
Then $H$ is strictly increasing, satisfies $(\mathrm{NPB})$, and
\[
 x+1\le H(x)\le G(x),\qquad
 \frac{H(x)-x}{G(x)-x}\longrightarrow0.
\]
If $G$ satisfies~\eqref{eq:sp}, so does $H$. If $G$ is primitive
recursive, so is $H$.
\end{theorem}
\begin{proof}
The positive surplus $\Delta$ is nondecreasing. Hence
$1\le\eta(x)\le\Delta(x)$ and $\eta$ is nondecreasing, proving
the order assertions. Put $Z_n=H^{[n]}(0)$ and $T_t=t(t+1)/2$.
At $Y_i$,
$\eta(Y_i)\ge\lceil(Y_{i+1}-Y_i)/(i+1)\rceil$; at most $i+1$
$H$-steps from any point at least $Y_i$ therefore reach $Y_{i+1}$.
Induction yields $Z_{T_t}\ge Y_t$. A polynomial bound on $Z_n$
would bound $Y_t$ polynomially, contradicting $(\mathrm{NPB})$.
Under~\eqref{eq:sp}, for $T_t\le n<T_{t+1}$,
$Z_n/n^k\ge Y_t/(t+2)^{2k}\to\infty$, proving the stronger claim.

Both $\Delta(x)$ and $d(x)$ tend to infinity: a bounded surplus
would give a linear orbit, and depth has the threshold identity
in Lemma~\ref{lem:geometry}. Given $M\ge1$, choose $a_M$ with
$d(u)+1\ge M$ for $u\ge a_M$. Splitting the maximum at $a_M$ gives
$\eta(x)\le\max(C_M,\Delta(x)/M+1)$ for a fixed $C_M$.
Thus $\limsup\eta/\Delta\le1/M$, and the ratio tends to zero.
For primitive-recursive $G$, its inverse is found by search bounded
by the input, depth by at most that many descents, and the displayed
finite maximum is primitive recursive.
\end{proof}
This construction is external; it does not claim that $H$ belongs
to an inverse-stride algebra. There is also no pointwise least
terminating generator preserving the noncollapse conclusions.
For every $M\ge0$,
\[
 G_M(x)=x+1+\max(x-M,0)
\]
agrees with successor through $M$ and has orbit
$Y_t=t$ for $t\le M$, $Y_{M+s}=M-1+2^s$ for $s\ge0$.
A pointwise least candidate would lie below all $G_M$ and above
successor, hence equal successor, whose stride families collapse.

\paragraph{Concrete spectra.}
For $G(x)=2x+1$,
$G^{[r]}(z)=2^r(z+1)-1$ and $\rho_r(y)=\lfloor y/2^r\rfloor$.
These equations, the minimal basis, and the earlier-majorant rule
are exactly the defining data of $\mathcal H^0_{1,r}$ in
\cite{Osipov2026}; thus $\B_{G,r}=\mathcal H^0_{1,r}$.
Reverse-divisibility inclusion already occurs in the canonical
lower-basis sectors of the earlier three-parameter Step Recursion
article \cite{Osipov2026}. The present doubling specialization
provides an explicit exact-spectrum and strict-intersection witness.
Our binary witness becomes
\[
 F_{2,3}(x,y)=
 \begin{cases}
 \lfloor x/4\rfloor,&(x,y)=(2^t-1,2^{t+1}-1)\text{ for some }t\ge0,\\
 0,&\text{otherwise}.
 \end{cases}
\]
On this support $\lfloor x/4\rfloor=\lfloor y/8\rfloor$, and both
membership spectra are $\{1,2,3\}$. This is inverse-depth geometry,
not a radix-specific phenomenon: for $G(x)=(x+1)^2$ the orbit begins
$0,1,4,25,676,458329$, and the same witness gives
$F_{2,3}(676,458329)=4$ with the same exact spectra.

\section{Conclusions and remaining questions}
The classification separates two kinds of intersection: indexed
single-stride algebras can leave the indexed family under meets,
while realized membership spectra are closed under finite intersections
in both recursion regimes. Shared bounding coordinates establish the
bounded closure without an unrestricted selector. Mixed strides
further realize finite monotone Boolean conditions on their additive
monoid; this is a realization theorem for fixed functions, not a
classification of all mixed-algebra inclusions.

Ordinary predecessor, doubling, and addition detect the polynomial
orbit threshold in the unbounded system. Bounded predecessor also
detects bounded generator jumps. Under those two conditions $\E$
embeds, and equality with $\E$ follows if the inverse belongs to $\E$.
A general criterion for full bounded collapse without that
inverse-membership assumption remains unresolved. For arbitrary $G$ under (PB), the
unbounded classes remain relative; their equality is equivalent to
$\rho_1\in\mathrm{PR}[\rho_r]$ for every $r$.

Further questions concern descriptions of nonprincipal functional
intersections, classification of unary and infinite spectra, finite-union
closure of arbitrary spectra, and mixed-algebra inclusion beyond the
finite Boolean realization. The general intersection theorem settles
the corresponding finite-meet closure problem for bounded spectra.

\appendix
\section{Basis-two complexity comparison}\label{app:basis-two}
For this comparison, define $\mathcal H^2_{1,1}$ explicitly as the
finite-stage closure of zero, successor, projections, addition
$e_1(x,y)=x+y$, and $e_2(x)=x^2+2$ under composition and bounded
step recursion along $\rho(y)=\lfloor y/2\rfloor$, with earlier
pointwise majorants. This is the \emph{basis-two} doubling-row
algebra of \cite{Osipov2026}. For $G(x)=2x+1$, the minimal-basis algebra
$\B_{G,1}=\mathcal H^0_{1,1}$ has only zero, successor, and
projections in its initial basis.
The following lemma proves polynomial-time containment directly.
Combined with the classical SAT representation in $\E$, it yields
the conditional incomparability below. This consequence is not
claimed as a new unconditional complexity-theoretic separation.

\begin{lemma}[Binary polynomial time at basis two]
\label{lem:basis-two-time}
Every fixed-arity function in $\mathcal H^2_{1,1}$ is computable
in polynomial time in the total binary input length.
\end{lemma}
\begin{proof}
For each fixed finite derivation $\delta$, we prove simultaneously
that there is a polynomial $Q_\delta\in\N[X]$ bounding every
numerical occurrence on input maximum $N$ by $Q_\delta(N+1)$,
and that its direct evaluation takes polynomial time in
$\ell=\lceil\log_2(N+2)\rceil$. Since arity is fixed, this
parameter is equivalent, up to constant factors, to total binary
input length. The basis functions have both properties, using
ordinary binary arithmetic. Composition preserves them: substitute
the constituent polynomial value bounds, and observe that all
head inputs then have binary length $O(\ell)$.

For a bounded recursion with earlier data $a,b,M$, choose the
polynomial occurrence bounds supplied by induction for these
three functions. On original input maximum $N$, every recursion
address is at most $N$. At every address $w$ of this call, the
genuine accumulator is at most $M(\mathbf x,w)$, hence at most
$Q_M(N+1)$. Thus the base and every transition have input maximum
at most $N+Q_M(N+1)$, a fixed polynomial in $N+1$. Applying
their inductive occurrence bounds gives a polynomial bound for
all internal occurrences as well. There are at most
$\lfloor\log_2 N\rfloor+1$ updates when $N>0$, and none when
the recursion address is zero. Every base and transition call has
binary input length $O(\ell)$, so its inductive evaluation time
is polynomial in $\ell$. Multiplying by the number of updates,
and including the polynomial cost of forming the inverse-address
list and maintaining the accumulator, preserves polynomial time.
The majorant is an admission condition and need not be evaluated;
evaluating it would also take polynomial time by induction.
This proves both assertions by structural induction.
\end{proof}

\begin{proposition}[A conditional incomparability at basis two]
\label{prop:e0-h2}
With the customary binary-input complexity convention,
\[
 \mathcal H^2_{1,1}\nsubseteq\E,
 \qquad
 \E\subseteq\mathcal H^2_{1,1}\ \Longrightarrow\
 \mathsf P=\mathsf{NP}.
\]
Consequently, if $\mathsf P\ne\mathsf{NP}$ then
$\E$ and $\mathcal H^2_{1,1}$ are incomparable.
The first noninclusion is unconditional. The incomparability
statement is conditional and does not settle whether
$\E\subseteq\mathcal H^2_{1,1}$.
\end{proposition}
\begin{proof}
Addition is an initial generator of $\mathcal H^2_{1,1}$,
whereas the shifted-coordinate envelope for ordinary bounded
recursion (Lemma~\ref{lem:envelope}) excludes addition from $\E$.
Hence $\mathcal H^2_{1,1}\nsubseteq\E$.
Lemma~\ref{lem:basis-two-time} gives
$\mathcal H^2_{1,1}\subseteq\mathsf{FP}$.
Grozea's construction \cite{Grozea2004} supplies a
fixed-arity SAT predicate in $\E$. Its inputs encode a CNF formula
as a packed symbol array, a symbol-width mask, and an upper bound
on assignments. After renumbering the variables, these parameters
can be produced in polynomial time with total polynomial binary
length; in particular, the assignment bound $2^n-1$ requires only
$n$ binary digits for $n$ variables. Thus a polynomial-time
algorithm for this predicate decides ordinary SAT in polynomial
time. If $\E\subseteq\mathcal H^2_{1,1}$, the predicate belongs
to $\mathsf{FP}$ by the lemma, and $\mathsf P=\mathsf{NP}$ follows.
\end{proof}

\section*{Declaration of generative AI and AI-assisted technologies in the manuscript preparation process}
During the preparation of this work, the author used ChatGPT (OpenAI) for
copyediting, grammatical correction, and language polishing of the manuscript.
The author reviewed and edited the resulting text as needed and takes full
responsibility for the content of the article.

\begingroup\small\raggedright

\endgroup
\end{document}